\documentclass[12pt]{article}
\usepackage{ascmac}
\usepackage{latexsym}
\usepackage[dvipdfmx]{graphicx}
\usepackage{amsmath}
\usepackage[top=3.5cm,bottom=3.5cm,left=2cm,right=2cm]{geometry}
\usepackage{pifont}          
\usepackage{amsmath,amssymb}
\usepackage{color}
\usepackage{geometry}
\usepackage{caption}
\usepackage{ulem}
\newcommand{\h}{\hspace}
\newcommand{\aln}[1]{\begin{align}#1\end{align}}

\begin{document}

\title{
\vbox{
\baselineskip 14pt
\hfill \hbox{\normalsize KUNS-2605
}} \vskip 1cm
\bf \Large  
Models of LHC Diphoton Excesses \\
Valid up to the Planck scale
\vskip 0.5cm
}
\author{
Yuta~Hamada\thanks{E-mail:  \tt hamada@gauge.scphys.kyoto-u.ac.jp},
Hikaru~Kawai\thanks{E-mail:  \tt hkawai@gauge.scphys.kyoto-u.ac.jp},
Kiyoharu~Kawana\thanks{E-mail: \tt kiyokawa@gauge.scphys.kyoto-u.ac.jp} and 
Koji~Tsumura\thanks{E-mail: \tt ko2@gauge.scphys.kyoto-u.ac.jp}
\bigskip\\
\it \normalsize
 Department of Physics, Kyoto University, Kyoto 606-8502, Japan\\
\smallskip
}
\date{\today}

\maketitle

\abstract{\normalsize
We discuss a possibility to explain the LHC diphoton excesses at $750$GeV by the new scalar $X$ that couples to the gauge bosons through the loop of new massive particles with Standard Model charges. We assume that the new particles decay into the Standard Model particles at the tree level. We systematically examine the models that preserve the vacuum stability and the perturbativity up to the Planck scale. When we take scalars for the new particles, we find that only a few diquark and dilepton models can explain the observed diphoton cross section without conflicting the experimental mass bounds. 
When we take vector-like fermions for the new particles, we find rather different situations depending on whether their couplings to $X$ are   
scalar or pseudoscalar type. In the former case, a few models are allowed if we introduce only one species of fermions. The more fermions  we introduce, the more models are allowed. In the later case, the most of the models are allowed because of the large coupling between $X$ and photon. It is interesting that the allowed mass regions of the scalar particles might be reached by the next lepton colliders.
}
\newpage

\section{Introduction}
%
The Standard Model (SM) for elementary particles has been completed by the Higgs boson discovery. 
The observed properties of the Higgs boson is totally consistent with that in the SM, and 
no direct evidence of the new physics beyond the SM has been found at the Large Hadron Collider (LHC) so far. 
On the other hand, we know several indirect signatures of the beyond the SM, e.g., 
the existence of dark matter, smallness of the neutrino mass and the baryon asymmetry of the universe. 
Furthermore, there are many open questions in the SM as a consistent theory such as 
the naturalness problem of the Higgs boson mass \cite{MPP,RGE}, that of the cosmological constant \cite{MEP}, 
the origin of the electroweak symmetry breaking \cite{CCS}, and so on. 
To solve these questions, various models have been proposed. 
Each of them has its own theoretical motivations and characteristic phenomenological predictions.  
In order to disentangle these models, 
some experimental inputs of new physics is quite important. \\

Recently, both the ATLAS and CMS collaborations have reported the excesses of the signal events 
in the diphoton invariant mass distribution at around $750$~GeV based on the first results of the LHC Run 2 data~\cite{ATLAS,CMS}. 
Although these signal significances are around $3\sigma$ confidence level, 
the results might be an indication of new physics beyond the SM. 
It would be valuable to study the interpretation of the observed excesses by simple scenarios 
because such a scenario can be understood as a low energy effective theory of more fundamental theories. 
%
By taking the Laudau-Yang theorem \cite{Landau:1948kw,Yang:1950rg} into the consideration, 
a possible simple scenario is that the new particle $X$ with $M_X\sim750$~GeV is a spin-zero SM singlet real 
scalar (or pseudoscalar) boson which couples to new massive scalar bosons or fermions charged under the SM gauge group \cite{Diphoton:X}. 
\footnote{
There can be other possible interpretations of the diphoton excesses. 
An interesting interpretation is that the peak in the diphoton is not a resonance of the new particle $X$ 
but a cusp in the loop integral of the box diagram of new particle $Y$ with $M_Y^{}\sim 375$~GeV \cite{Diphoton:Y}.
}
\\

\begin{figure}[tbh]
\centering
\includegraphics[width=6cm]{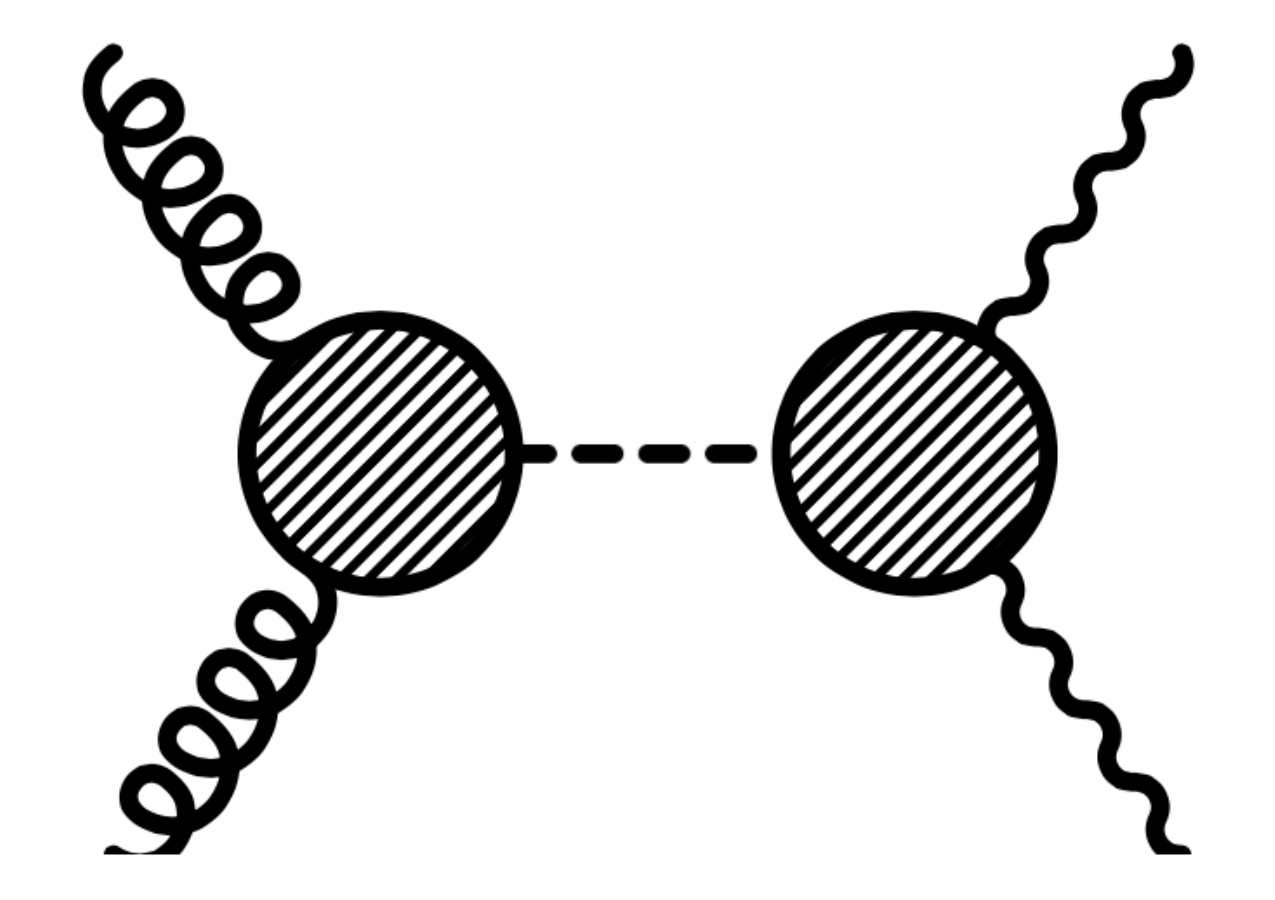}
\caption{A event topology of the simple interpretation for the LHC diphoton excess. 
The new particle $X$ with $M_X=750$ is produced by the gluon-fusion, and decays into a pair of photons.}
\label{FIG:diagram}
\end{figure}

Before going into details of the models, let us summarize the general aspects of $X$. 
When $X$ is a scalar, its effective couplings to the SM gauge bosons are parametrized as 
\begin{equation}  {\cal{L}}_{\text{ eff}}\ni -\frac{X}{4}\left(g_{X\gamma}^{}F^{\mu\nu}F_{\mu\nu}+g_{Xg}^{}G^{\mu\nu}G_{\mu\nu}\right), \end{equation} 
where the dimensionful parameters $g_{X\gamma}^{}$ and $g_{Xg}^{}$ are determined 
by evaluating loop diagrams when we fix the model. 
Similarly, when $X$ is a pseudoscalar, the effective Lagrangian is given by 
\begin{equation}   {\cal{L}}_{\text{ eff}}\ni-\frac{X}{4}\left(g_{X\gamma}^{}F^{\mu\nu}\tilde{F}_{\mu\nu}+g_{Xg}^{}G^{\mu\nu}\tilde{G}_{\mu\nu}\right).\end{equation} 
Here, $F^{\mu\nu}$ and $G^{\mu\nu}$ are the field strengths of the $U(1)_{\text{em}}$ and $SU(3)_{c}$ respectively. 
Using these dimension-five interactions, the diphoton signal at the LHC can be interpreted 
as the process $gg \to X \to \gamma\gamma$ in Fig.~\ref{FIG:diagram}. 
The diphoton cross section $\sigma_{\gamma\gamma}^{}$ is easily estimated 
by the gluon-fusion production cross-section times 
$\gamma\gamma$ branching ratio of $X$ at 13TeV. 
In this paper, we follow the result in \cite{Ellis:2015oso}:
\aln{ \sigma_{\gamma\gamma}:&=\sigma(gg\rightarrow X)\times \text{Br}(X\rightarrow\gamma\gamma)\nonumber\\
&=(13\text{pb})\times(g_{X\gamma}\cdot \text{TeV})^2. \label{eq:theoretical cs}}
Note that the main decay mode of $X$ is assumed to be $gg$ channel
\footnote{This assumption typically predicts the narrow decay width of $X$ which is favored by CMS data \cite{CMS}. 
On the other hand, ATLAS results prefer the wide width of $\sim 45$ GeV \cite{ATLAS}.}. 
Thus, the dependence of $g_{Xg}^{}$ in the overall reaction rate is almost canceled out. 
Throughout this paper, we assume the mass of internal particles in the loop 
to be heavier than the half of $M_X$ in order to forbid the tree level decay of $X$. 
On the other hand, the experimental data implies \cite{Ellis:2015oso}
\begin{equation}  \sigma_{\gamma\gamma}=1\text{--}10~\text{fb}.\label{eq:observed cs}\end{equation} 
By calculating the effective coupling $g_{X\gamma}^{}$ in a specific model, 
we can compare Eq.(\ref{eq:theoretical cs}) with Eq.(\ref{eq:observed cs}) and 
discuss whether such a model is favorable in the context of the experimental data.
\\

The coupling $g_{X\gamma}^{}$ is generated at one-loop in each simple scenario.
Thus, its magnitude is typically small due to not only the one-loop suppression but also the new mass scale. 
In order to generate sufficiently large $g_{X\gamma}^{}$, 
we need relatively strong interactions among $X$ and the new SM charged particles. 
It is known that such strong interactions often violate theoretical consistencies of the model. 
For instance, a large trilinear scalar coupling $\mu$ 
tends to induce the vacuum instability. In order to rescue the vacuum stability, one may introduce 
a large scalar quartic coupling $\lambda_{\phi}^{}$. However, such $\lambda_{\phi}^{}$ leads 
a rapid blowup of the running coupling constant and breaks the perturbativity below the Planck scale. 
Therefore, requirements of theoretical consistency give non-trivial constraints
on the models of the diphoton excesses. \\

In this paper, we study whether simple scenarios can explain the diphoton excesses at the LHC  
while preserving the vacuum stability of the scalar potential and the perturbativity of the gauge, Yukawa and 
scalar quartic coupling constants up to the Planck scale 
\footnote{
For the scenarios with the vector-like fermions, the similar analysis was first done in \cite{Son:2015vfl} where they discuss the perturbativity and the vacuum stability of the models. Our analysis is more inclusive in that we also consider the extensions with the new SM charged scalars, and include the new scalar couplings. These couplings play important role to determine the Landau pole. See Section \ref{sec:scalar} for the details.
}
. 
As for the new SM charged particles, we consider scalars (diquark, leptoquark and dilepton) or fermions 
(vector-like quark and vector-like lepton) which can decay to the SM particles at tree level.
For the completeness, we also consider the cases with $N_{f}$ multiplets of the new SM charged particles 
in order to maximize the diphoton cross sections. 
To study the perturbativity up to the Planck scale, we use the one-loop renormalization group equations (RGEs)
.
If we consider a scalar $(N_f=1)$ as the new particles, only a few models can explain the observed cross section  
without conflicting the experimental mass bounds. The allowed models require diquarks and dileptons near the threshold masses ($\lesssim 450$GeV). 
All the leptoquark models are disfavored in our analysis 
due to the strong lower mass bounds at the LHC and the rapid running of the scalar quartic coupling constants. 
The increase of $N_f$ makes the situation better, however, too large $N_f$ is disfavored by the RGE of 
the quartic coupling of the new scalars. 
On the other hand, if we take vector-like fermions as the new particles, the effective coupling $g_{X\gamma\gamma}$ 
behaves differently depending on the type of the Yukawa couplings between $X$ and the fermions. 
In the case of the scalar type coupling, only a few models are allowed if $N_{f}=1$. 
If we increase $N_{f}$, more models are allowed. 
Furthermore, in the case of the pseudoscalar type coupling, almost all of the models are allowed 
because of the enhancement of the effective coupling $g_{X\gamma\gamma}$ due to the property of the one-loop integral. 
 \\

This paper is organized as follows. 
In Section \ref{sec:scalar}, we study the SM extended by $X$ and the new SM charged scalars.
Stability and perturbativity bounds of the new scalar sector are discussed 
as well as the interpretation of the LHC diphoton excess. 
In Section \ref{sec:fermion}, the scenarios with new vector-like fermions are considered in a similar manner. 
Conclusions and discussion are given in Section \ref{sec:summary}. In \ref{app:bounds}, we summarize the direct search mass bounds for the new particles. In \ref{app:rge}, the one loop RGEs are presented.

\section{Extensions with New Scalar Bosons}\label{sec:scalar}
In this section, we study the extensions of the SM by new scalar multiplets with SM charges, which 
will be regarded as the source of $g_{X\gamma}^{}$ 
via the one-loop diagram. 
Requiring tree-level decays of the new scalar bosons,\footnote{
The abundance of long-lived charged particles is strongly constrained by cosmology ~\cite{Kang:2006yd}. On the other hand, color singlet particles are allowed as a part of Dark Matter \cite{MDM}. For those extensions, the following discussion can be equally applied. 

}
their quantum charges of the new scalar fields are determined by the combination of the SM fermions. 
We list all the possible charge assignments of the new scalar multiplets 
in Table~\ref{Tab:DQs} for diquarks, Table~\ref{Tab:LQs} for leptoquarks, and Table~\ref{Tab:DLs} for dileptons.  
Note that we include right-handed neutrinos $\nu_{R}^{}$ as SM fermions. 
In these Tables, the normalization of the hypercharge is given by $I_3+Y_{\phi}=Q$. \\

\begin{table}[t]
\begin{center} 
\begin{tabular}{c||c|c|c||c|c}
 {}
 & $SU(3)_c$ & $SU(2)_L$ & $U(1)_Y$ & $\lambda_{\phi}^{\text{Max}}$&
 $(N_f^2 \lambda_\phi)^\text{Max}$  \\\hline\hline
$DQ_0^d$ & ${\bf 3\,\, (6^*)}$ & ${\bf 1}$ & $-1/3$  &$0.32\,\,(\times)$
                  &$2.2\,|_{N_f=6}$\\\hline
$DQ_0^y$ & ${\bf 3\,\, (6^*)}$ & ${\bf 1}$ & $-4/3$  &$0.34\,\,(\times)$
                  &$1.2\,|_{N_f=3}$\\\hline
$DQ_0^u$ & ${\bf 3\,\, (6^*)}$ & ${\bf 1}$ & $2/3$  &$0.33\,\,(\times)$
                  &$2.2\,|_{N_f=6}$\\\hline
$DQ_1$ & ${\bf 3\,\, (6^*)}$ & ${\bf 3}$ & $-1/3$  &$0.13\,\,(\times)$& $0.13\,|_{N_f=1}$\\\hline
$DQ_{1/2}$ & ${\bf 1_H\,\, (8)}$ & ${\bf 2}$ & $1/2$ &$0.18\,\,(\times)$
                    &$8.5\,|_{N_f=20}$ 
\end{tabular}
\caption{A list of diquarks which couple to the SM quarks. Here, $(N_{f}^2\lambda_{\phi})^\text{Max}$  is the maximal value of $N_{f}^2\lambda_{\phi}^\text{Max}$ where $\lambda_{\phi}^\text{Max}$ is calculated for each $N_f$ by requiring the perturbativity up to the Planck scale.
}
\label{Tab:DQs}
\end{center}
\end{table}

\begin{table}[t]
\centering
\begin{tabular}{c||c|c|c||c|c|c}
 {}
 & $SU(3)_c$ & $SU(2)_L$ & $U(1)_Y$  & $\lambda_{\phi}^\text{Max}$
 &$(N_f^2 \lambda_\phi)^\text{Max}$ \\\hline\hline
$S_0^{d*}$ & ${\bf 3^*}$ & ${\bf 1}$ & $1/3$ &$0.32$
                   &$2.2\,|_{N_f=6}$ \\\hline
$S_0^{y*}$ & ${\bf 3^*}$ & ${\bf 1}$ & $4/3$ &$0.34$ 
                   &$1.2\,|_{N_f=3}$ \\\hline
$S_1^*$     & ${\bf 3^*}$ & ${\bf 3}$ & $1/3$ &$0.13$ &$0.13\,|_{N_f=1}$ \\\hline
$S_{1/2}$   & ${\bf 3}$ & ${\bf 2}$ & $7/6$  &$0.25$
                  &$0.46\,|_{N_f=2}$ \\\hline
$S^q_{1/2}$ & ${\bf 3}$ & ${\bf 2}$ & $1/6$ &$0.24$
                   &$0.59\,|_{N_f=3}$  \\ \hline 
$R_0^{u*}$ & ${\bf 3^*}$ & ${\bf 1}$ & $-2/3$ &$0.33$
                   &$2.2\,|_{N_f=6}$\\\hline
$R_0^{d*}$ & ${\bf 3^*}$ & ${\bf 1}$ & $1/3$ &$0.32$
                   &$2.2\,|_{N_f=6}$ \\\hline
$R_{1/2}^*$ & ${\bf 3}$ & ${\bf 2}$ & $-1/6$ &$0.24$
                    &$0.59\,|_{N_f=3}$
\end{tabular}
\caption{A list of leptoquarks which couple to the SM fermions.
}
\label{Tab:LQs}
\end{table}

\begin{table}[tbh]
\centering
\begin{tabular}{c||c|c|c||c|c|c}
 {}
 & $SU(3)_c$ & $SU(2)_L$ & $U(1)_Y$ & $\lambda_{\phi}^\text{Max}$
 &$(N_f^2 \lambda_\phi)^\text{Max}$ \\\hline\hline
$h_0^+$ & ${\bf 1}$ & ${\bf 1}$ & $1$  &$0.25$
                                 &$19\,|_{N_f=25}$ \\\hline
$h_0^{++}$ & ${\bf 1}$ & ${\bf 1}$ & $2$  &$0.27$
                                 &$1.1\,|_{N_f=3}$ \\\hline
$\Delta_1$ & ${\bf 1}$ & ${\bf 3}$ & $1$  &$0.27$
                                  &$0.77\,|_{N_f=3}$ \\\hline
$\Phi_{1/2}$ & ${\bf 1_H}$ & ${\bf 2}$ & $1/2$ &$0.25$ 
                                  &$8.5\,|_{N_f=20}$  \\\hline 
$s_0^{}$ & ${\bf 1}$ & ${\bf 1}$ & $0$  & $\times$&$\times$
\end{tabular}
\caption{A list of dileptons which couple to leptons. 
}
\label{Tab:DLs}
\end{table}

The diquark field in Table~\ref{Tab:DQs} is {\bf 3} or {\bf 6*} representation under the QCD as $(q\overline{q^c})$ states 
and {\bf 1} or {\bf 8} representation as $(q\bar q)$ states. 
The {\bf 1} representation is nothing but a Higgs doublet in the SM. 
Except for color charge assignments, there are five variations of diquarks. 
Each of them has specific decay pattern depending on the Yukawa couplings (see \ref{app:bounds}). 
%
Eight possible leptoquark fields are listed in Table~\ref{Tab:LQs}.
All of them are {\bf 3} representation under the color gauge group since they are $(q\overline{\ell^c})$ or $(q\bar\ell)$ states. 
In addition to the conventional scalar leptoquarks listed by PDG \cite{PDG}, 
we here include $R_0^u$, $R_0^d$ and $R_{1/2}^{}$ leptoquarks since we regard $\nu_R^{}$ as the SM fermions. 
%
All possible dilepton fields are given in Table~\ref{Tab:DLs}. 
In this class of extended models, we need to introduce additional colored particles $S_c$ in order to guarantee 
the sufficiently large production cross section of $gg\to X$.\footnote{
In principle, photon-fusion can be a source of $X$ production if we take photon into account as a parton \cite{PhotonPDF}. 
However, it is required to have extremely large $g_{X\gamma}^{}$ coupling at one-loop level. }
In the following, we neglect the contribution to $g_{X\gamma}^{}$ from $S_c$ for simplicity. 
This situation is realized if $S_c$ has a relatively small hypercharge. 
In \ref{app:bounds}, we give a list of conservative estimates for the lower mass bounds of the new scalar bosons, 
which will be compared with the bound derived from theoretical considerations. \\

%
%
The scalar potential of each model is generally given by
\aln{ V&
=
-\frac{M_{h}^2}{2}\left(H^{\dagger}H\right)
+M_{\phi}^2\left(\phi^{\dagger}\phi\right)
+\frac{M_{X}^2}{2}X^2
+\mu\left(\phi^\dagger\phi\right)X
+\mu'\left(H^\dagger H\right)X
+\frac{\mu_{X}^{}}{3!}X^3\nonumber\\
&+\lambda\left(H^{\dagger}H\right)^2
+\lambda_{\phi}\left(\phi^{\dagger}\phi\right)^2
+\frac{\lambda_{X}}{4!}X^4
+\lambda'_{\phi}\left(\phi^{\dagger}T_{\phi}^{A}\phi\right)^2
+\lambda''_{\phi}\left(\phi^{\dagger}t_{\phi}^{a}\phi\right)^2
+\lambda'''_{\phi}\left(\phi^{\dagger}T_{\phi}^{A}t_{\phi}^{a}\phi\right)^2\nonumber\\
&+\lambda_{\text{tr}}\left(\phi^{\dagger i}\phi_{j}\right)\left(\phi^{\dagger j}\phi_{i}\right)
+\lambda_{\phi 2}\left(\phi^\dagger\{t_{\phi}^a,t_{\phi}^b\}\phi\right)^2+\lambda_{\text{ad}}\text{Tr}\left(\tilde{\phi}_{\alpha}\tilde{\phi}_{\beta}\right)\text{Tr}\left(\tilde{\phi}^{\alpha\dagger}\tilde{\phi}^{\beta\dagger}\right)+\lambda_{\text{ad}}'\text{Tr}\left(\tilde{\phi}_{i}'\tilde{\phi}_{j}'\right)\text{Tr}\left(\tilde{\phi}'^{i\dagger}\tilde{\phi}'^{j\dagger}\right)
\nonumber\\
&+\kappa_{H\phi}^{} \left(H^{\dagger}H\right)\left(\phi^{\dagger}\phi\right)
+{\kappa'}_{H\phi}^{} \left(H^{\dagger}t_{H}^{a}H\right)\left(\phi^{\dagger}t_{\phi}^{a}\phi\right)
+\frac{\kappa_{HX}^{}}{2} \left(H^{\dagger}H\right)X^2+\frac{\kappa_{\phi X}^{}}{2} \left(\phi^{\dagger}\phi\right)X^2+\cdots,\label{eq:pot1}}
where $\phi$ is one of the scalar fields listed in Tables 1-3, $i(j)$ is the $SU(2)_{L}$ index of $\phi$, $H$ is the SM Higgs doublet field, $t^{a}_{S}\, (a=1$--$3)$ is the $SU(2)_{L}$ generator for $S=H, \phi$, 
$T_{\phi}^A\,(A=1$--$8)$ is the $SU(3)_{c}$ generator of $\phi$, 
and $\tilde\phi\h{1mm}(\tilde{\phi}'):=\sum_{a}\phi^at_{\phi}^a\h{1mm}(\sum_{A}\phi^AT^A)$ exists only when the representation of $\phi$ is the adjoint representation. 
In the following discussion, 
we put 
$\mu'=\mu_{X}^{}=\lambda'_{\phi}=\lambda''_{\phi}=\lambda'''_{\phi}=\lambda_{\text{tr}}
=\lambda_{\phi2}=\lambda_{\text{ad}}=\lambda_{\text{ad}}'
=\kappa_{H\phi}^{}={\kappa'}_{H\phi}^{}=\kappa_{\phi X}^{}=0$ 
at the weak scale for simplicity 
\footnote{
These couplings and a large number of flavors (particles) can potentially bring some conflicts with the electroweak precision test (EWPT) \cite{EWPT}. For example, the mixing coupling $\kappa_{H\phi}'$ generates the mass splitting of $\phi$, and contributes to the oblique parameters. 
Thanks to our simplified choice of parameters, the constraints from the EWPT are evaded. Furthermore, as for the cubic couplings, they can induce a non-zero vacuum expectation value (vev) of $X$. Such a vev can generally decrease $\sigma_{\gamma\gamma}$ because the total decay width of $X$ becomes larger through the possible direct decay $X\rightarrow \bar{H}H$. 
As for the scalar quartic couplings, there might be parameter regions such that $\sigma_{\gamma\gamma}^{}$ becomes larger. For example, when the mixing couplings such as $\kappa_{H\phi}$ are negative, the vacuum stability is not guaranteed any more, but we can rescue it by making other quartic self-couplings positively large. In principle, such choices can increase the upper bound of $\mu$ (see Eq.(\ref{eq:uppermu})). However, we must simultaneously consider the RGE effects of these couplings, and such effects can be stronger than the former. Seeking for some fine-tuned parameter regions is beyond the one-loop analysis of this paper. 
}. 

As we noted in Introduction, the diphoton decay of $X$ is induced at one-loop level. 
The effective coupling $g_{X\gamma}^{}$ is calculated as
\begin{equation}  
g_{X\gamma}^{}
=\frac{N_{f}\,\alpha \sum_{\phi} Q^2_{\phi}}{2\pi}\frac{\mu}{M_{X}^2}\, 
f(M_{\phi}^2/M_{X}^2),
\label{eq:effcoupling1}
\end{equation} 
where $\alpha(=e^2/4\pi)$ is the fine-structure constant, $\sum_{\phi} Q_\phi^2$ stands for summation of $U(1)_{\text{em}}$ charge squared over the multiplet $\phi$ including the color factor,
and $N_f$ denotes the number of flavors of $\phi$. 
Because we simplify the scalar potential by taking $\kappa_{H\phi}^{}={\kappa'}_{H\phi}^{}=0$, 
components of $\phi$ are degenerate in mass $M_\phi$. 
The one-loop integral is easily evaluated as 
\begin{equation}  
f(x):
=8x\left(\arctan\frac{1}{\sqrt{4x-1}}\right)^2-2.
\end{equation} 
The function takes the maximal value $f(1/4)=(\pi^2-4)/2$ at the threshold $M_X=2M_\phi$, 
and $\phi$ decouples monotonically for large $M_\phi$, i.e., $f(\infty)=0$.

In order to obtain the large diphoton cross section, the scalar trilinear coupling $\mu$ is required to be large as in Eq.~\eqref{eq:effcoupling1}. 
On the other hand, too large $\mu$ is constrained by the vacuum stability. 
By focusing on $ M_{X}^2$, $\mu$ and $ \lambda_{\phi}$ terms in Eq.(\ref{eq:pot1}), we have
\aln{ 
V&\ni 
\frac{M_{X}}{2}X^2
+\mu\left(\phi^\dagger\phi\right)X
+\lambda_{\phi}\left(\phi^\dagger\phi\right)^2\nonumber\\
&=\frac{M_{X}^2}{2}\left(X+\frac{\mu}{M_{X}^2}(\phi^\dagger\phi)\right)^2
+\left(\lambda_{\phi}-\frac{\mu^2}{2M_{X}^2}\right)(\phi^\dagger\phi)^2,
}
which leads to 
\begin{equation}  
\mu^2<2\lambda_{\phi}M_{X}^2\label{eq:uppermu}
\end{equation} 
by imposing the stability of the scalar potential at tree level
\footnote{If $X$ has a non-zero vacuum expectation value $\langle X\rangle$, the $(\phi^\dagger\phi) X$ coupling gets another contribution from $(\phi^\dagger\phi) X^2$. However, the total effective coupling $\tilde{\mu}=\mu+\langle X\rangle \kappa_{\phi X}/2$ also follows the same equation as Eq.(\ref{eq:uppermu}). Therefore, our examination Eq.(\ref{eq:maxcs1}) does not change. Moreover, non-zero $\langle X\rangle$ opens up new decay channels such as $X\rightarrow HH$ through the $(H^\dagger H)X^2$ coupling, which leads to the decrease of $\sigma_{\gamma\gamma}^{}$. Thus, to maximize $\sigma_{\gamma\gamma}$, we assume $\langle X\rangle=0$ in this paper.
}. 
Eq.~\eqref{eq:uppermu} shows that large $\lambda_\phi$ can help the vacuum stability against $\mu$.

Furthermore, the magnitude of $\lambda_{\phi}$ is also constrained by requiring the perturbativity up to the Planck scale. 
It is known that a large quartic coupling constant is a source of the Landau pole at high energy by considering 
the renormalization group evolution. 
In our setup, in order to ensure the substantial diphoton signal rate, small $\lambda_{\phi}$ is disfavored through the vacuum stability condition. Therefore, $\lambda_{\phi}$ tends to be non-perturbative at high energy.  
In Table~\ref{Tab:DQs}, \ref{Tab:LQs} and \ref{Tab:DLs}, 
we list $\lambda_{\phi}^\text{Max}$ (the maximal values of $\lambda_{\phi}$ at weak scale), 
which keeps the purturbativity up to the Planck scale, 
in various scalar extended models. The values are calculated using the one-loop RGEs in \ref{app:rge} \footnote{
In our calculation, the contributions from the possible Yukawa couplings are neglected. 
Indeed, the new particles decay promptly even for very small Yukawa couplings. On the other hand, if the Yukawa couplings are large, $\lambda_{\phi}^{\text{Max}}$ would be increased. 
}. 
On one hand, to maximize the diphoton signal rate, larger $N_f$ seems to be favorable. 
On the other hand, the perturbativity bounds become more severe. 
We determine $(N_f^2 \lambda_\phi)^\text{Max}$, the maximal value of $N_f^2 \lambda_\phi^\text{max}$ where 
$\lambda_{\phi}^{\text{Max}}$ is calculated for each $N_f$ by requiring the pertubativity condition up to the Planck scale. In the following, we denote the corresponding $N_f$ as $N_{\text{Max}}^{}$.
Notice that when the $SU(3)_{c}$ representation of ${\bf 6^*}$ and ${\bf 8}$, 
the Landau pole of $\lambda_{\phi}$ appears below the Planck scale even 
if $\lambda_{\phi}$ is set to be zero at the weak scale, see also Ref.~\cite{LP} for the $SU(2)$ case.

\begin{figure}
\begin{center}
\begin{tabular}{c}
\begin{minipage}{0.5\hsize}
\begin{center}
\includegraphics[width=8cm]{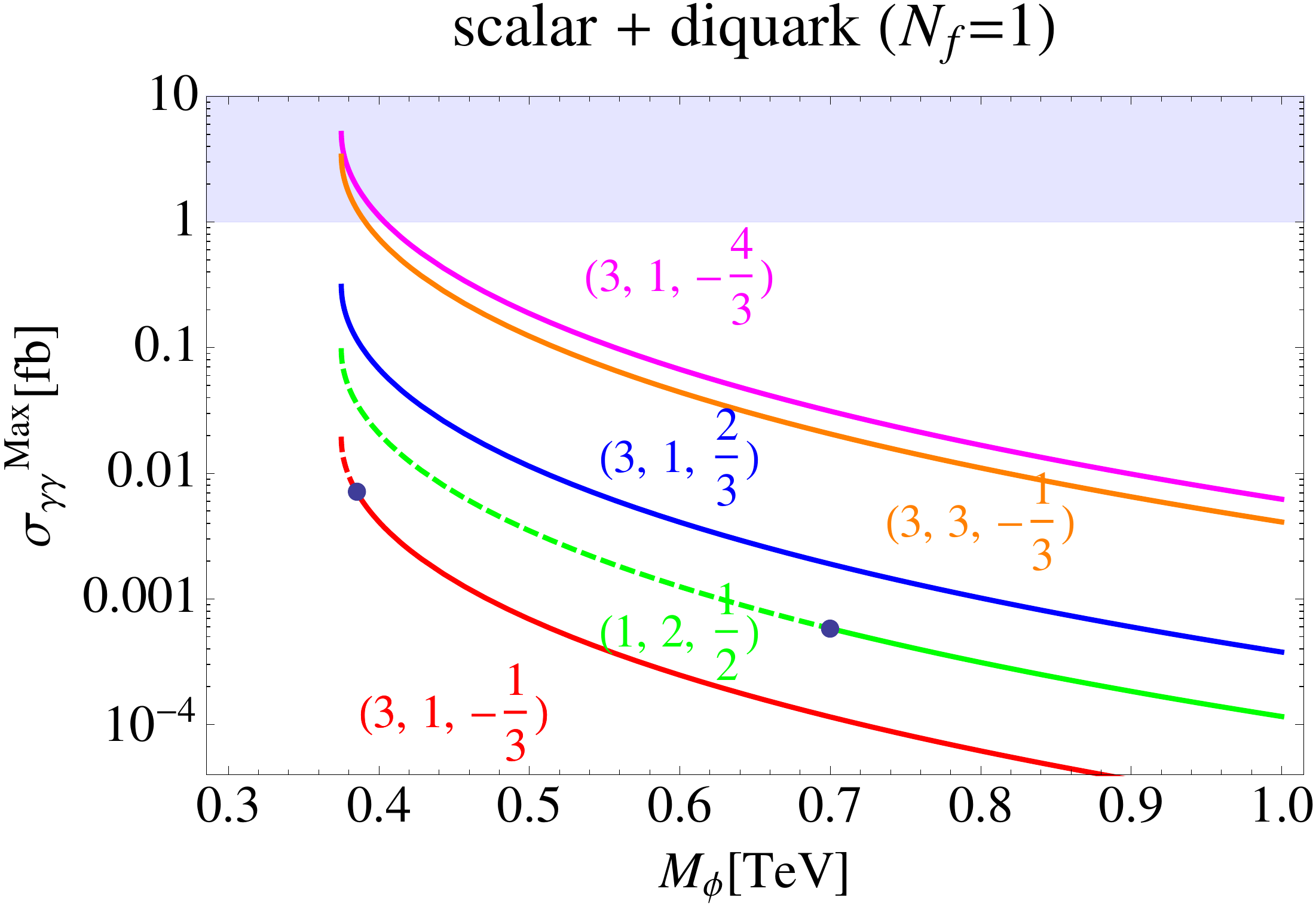}
\end{center}
\end{minipage}
\begin{minipage}{0.5\hsize}
\begin{center}
\includegraphics[width=8cm]{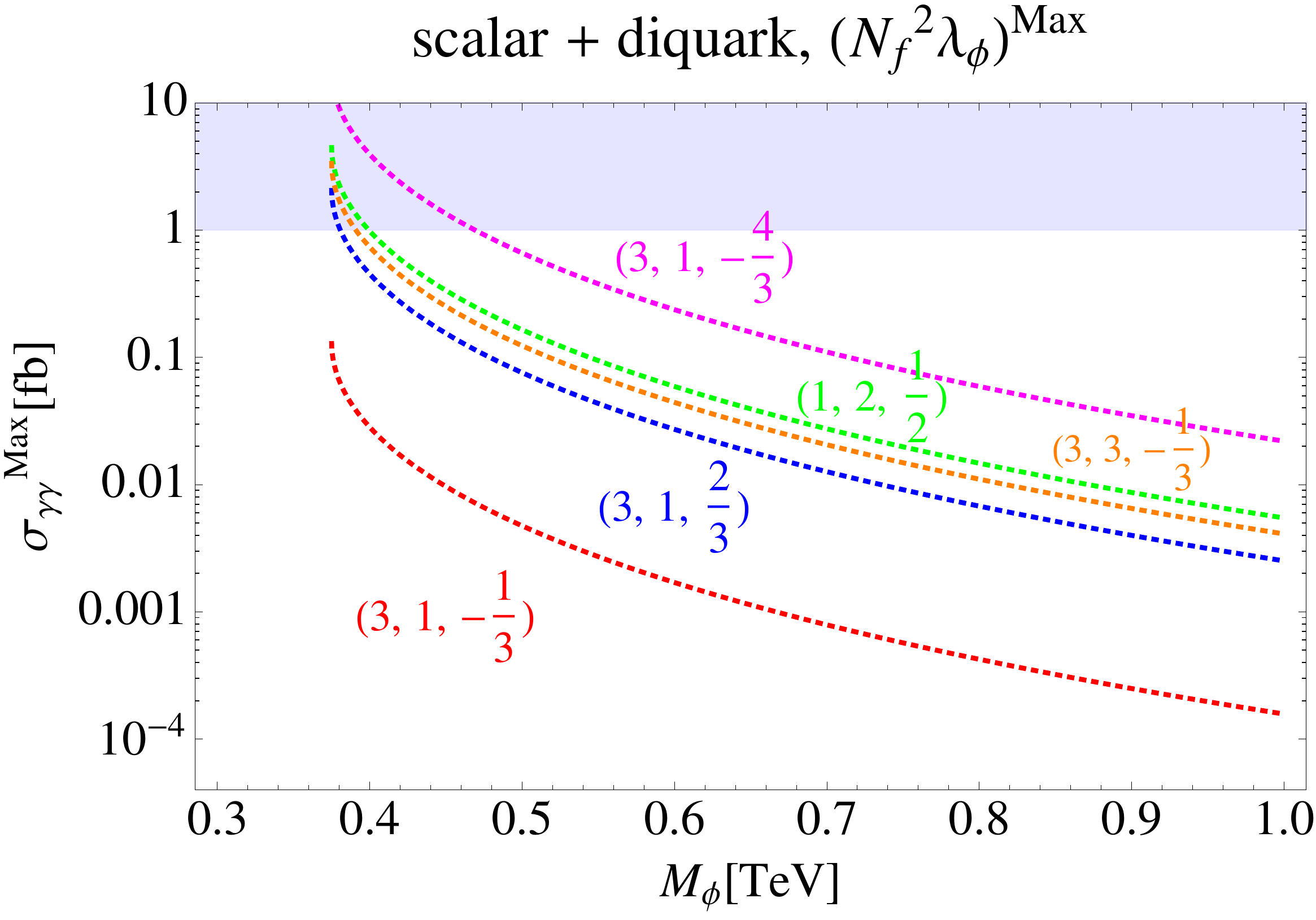}
\end{center}
\end{minipage}
\\
\\
\begin{minipage}{0.5\hsize}
\begin{center}
\includegraphics[width=8cm]{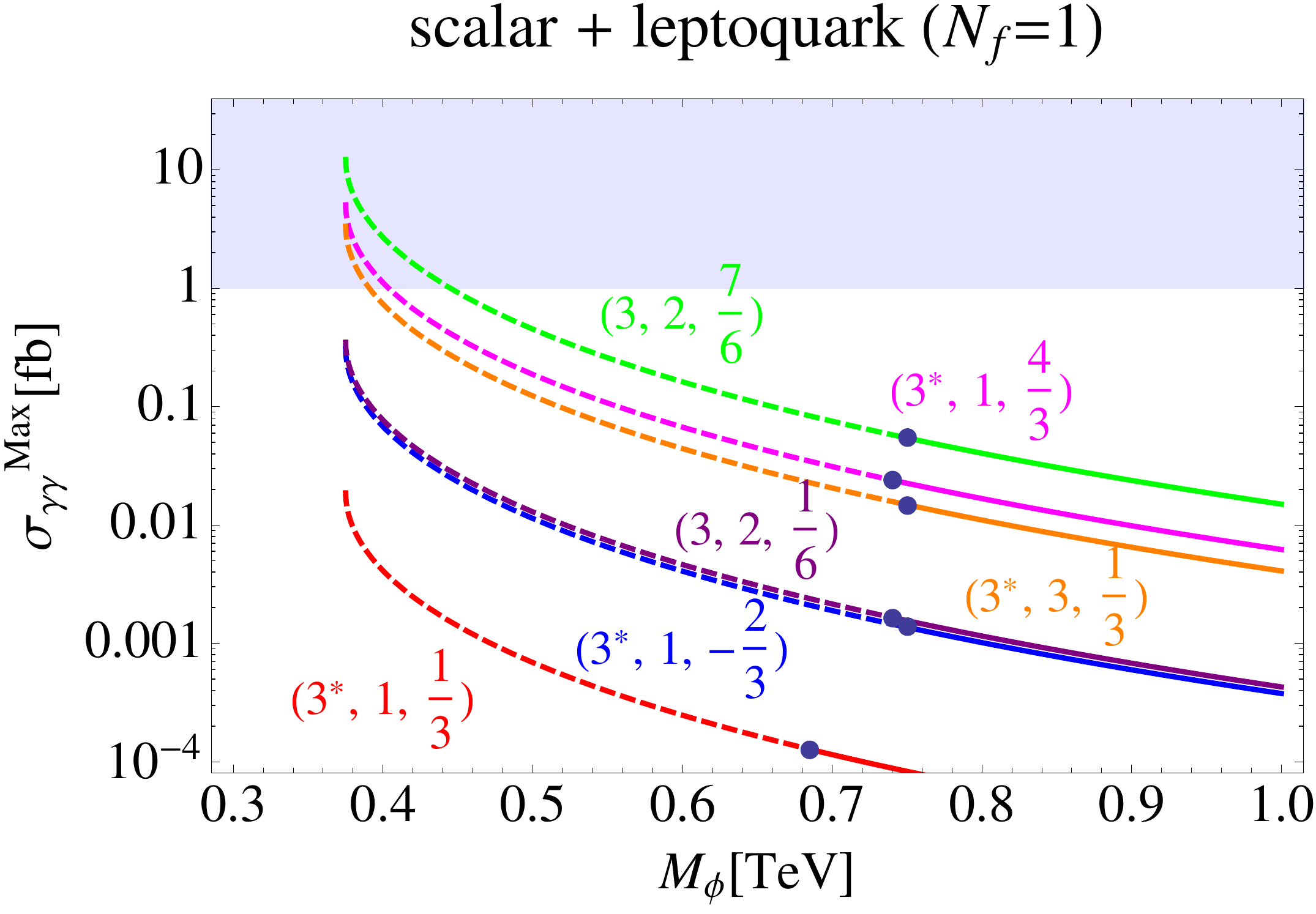}
\end{center}
\end{minipage}
\begin{minipage}{0.5\hsize}
\begin{center}
\includegraphics[width=8cm]{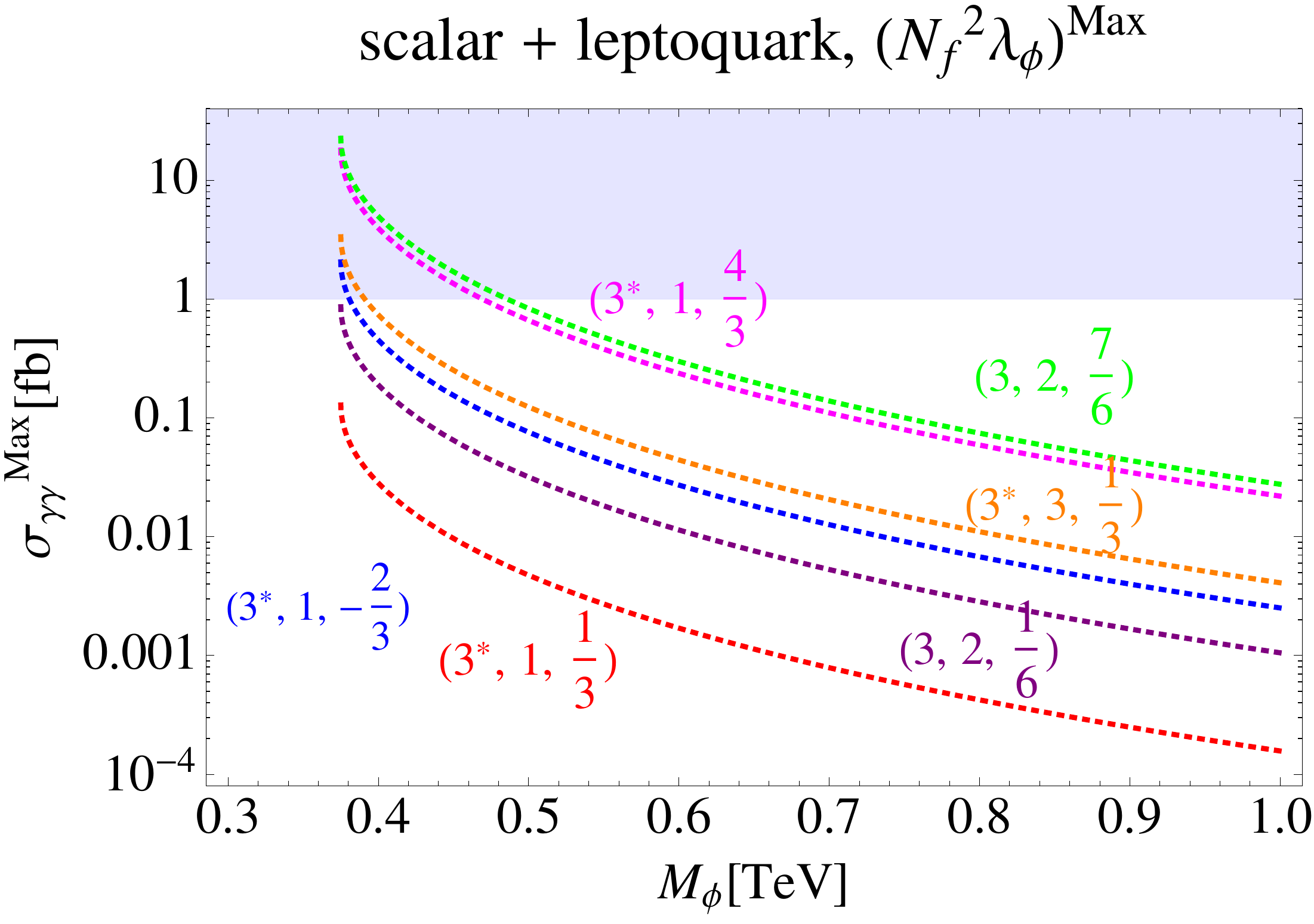}
\end{center}
\end{minipage}
\\
\\
\begin{minipage}{0.5\hsize}
\begin{center}
\includegraphics[width=8cm]{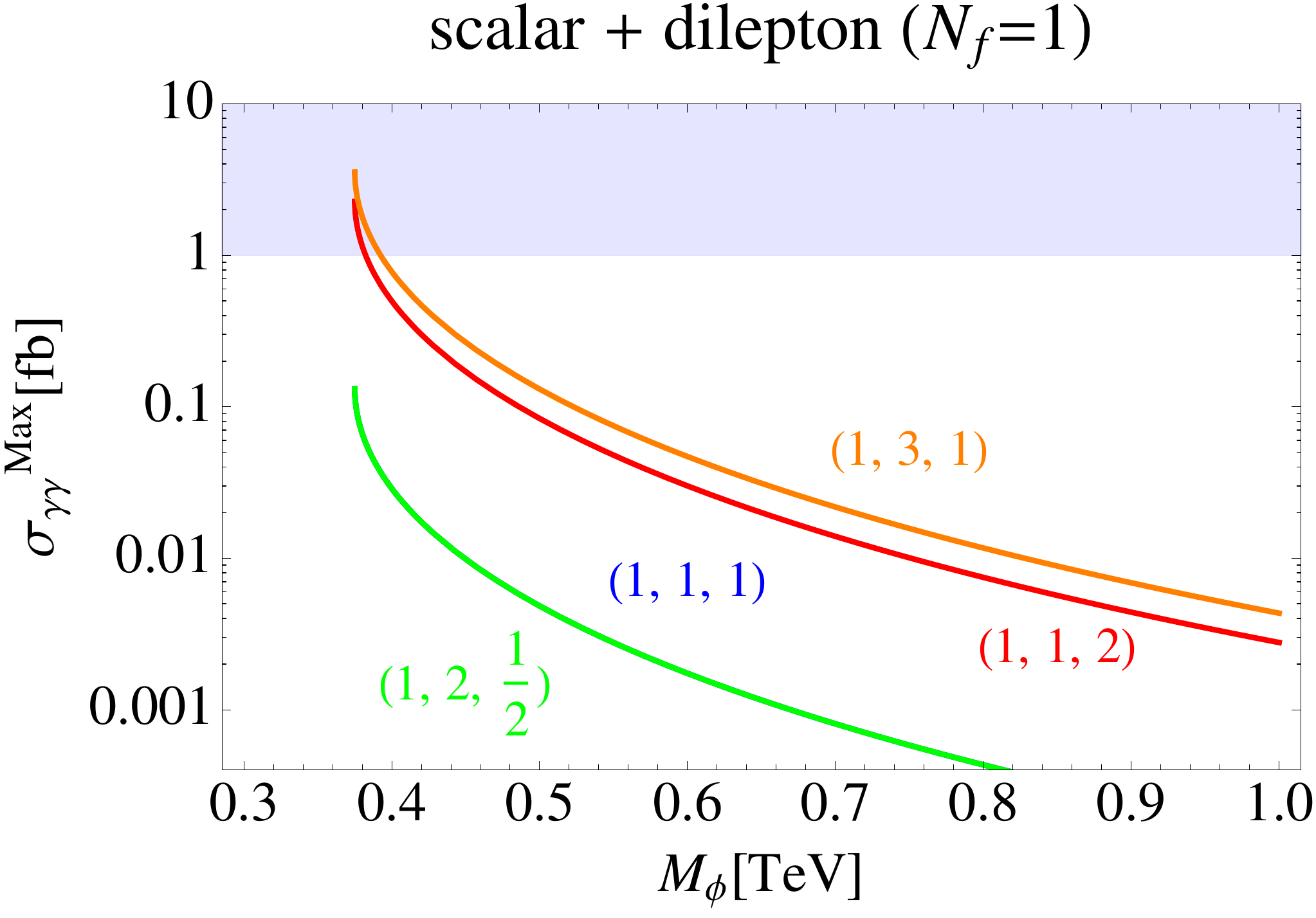}
\end{center}
\end{minipage}
\begin{minipage}{0.5\hsize}
\begin{center}
\includegraphics[width=8cm]{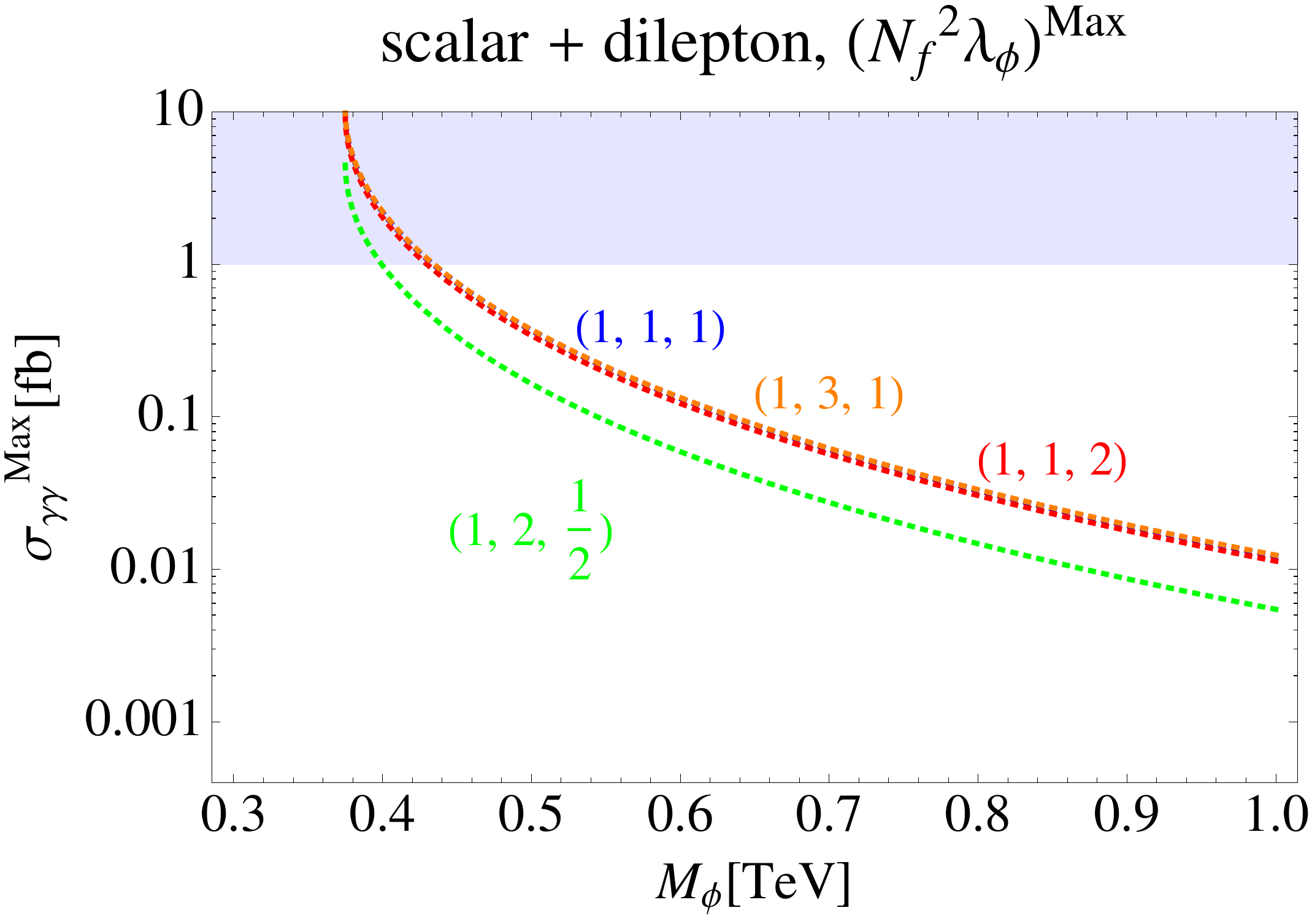}
\end{center}
\end{minipage}
\end{tabular}
\end{center}
\caption{
The maximal value, $\sigma_{\gamma\gamma}^\text{Max}$ for the production cross sections 
times $\gamma\gamma$ branching ratios, are shown as functions of $M_{\phi}$, the mass of the new SM charged scalar boson. 
Each curve shows the value calculated in the different scalar extension models. 
The solid (dashed) part of curves is the allowed (excluded) mass range by the collider experiments for $N_f=1$. 
For $N_f>1$, the results are given by dotted curves. 
The shaded region implies the required cross section from the\h{5mm} LHC diphoton excess, i.e., $\sigma_{\gamma\gamma}^\text{Max}>1$fb. 
}
\label{fig:scalar cs}
\end{figure}

Combining Eqs.(\ref{eq:theoretical cs}), (\ref{eq:effcoupling1}) and (\ref{eq:uppermu}), 
we calculate the maximum value of $\sigma_{\gamma\gamma}$:
\begin{equation}  
\sigma_{\gamma\gamma}^\text{Max}
=46\text{pb}\times \lambda_{\phi}^\text{Max}
\times\left(\frac{N_{f}\,\alpha \sum_{\phi}Q^2_{\phi}}{2\pi}\,f(M_{\phi}^2/M_{X}^2)\right)^2
\times\left(\frac{750\text{GeV}}{M_{X}}\right)^2.\label{eq:maxcs1}
\end{equation} 
In Fig.\ref{fig:scalar cs}, we plot the maximal $\sigma_{\gamma\gamma}^{}$ in Eq.(\ref{eq:maxcs1}) 
as functions of $M_{\phi}$ for various scalar extended models. 
The left (right) panels show the case with $N_f=1\, (N_\text{Max})$. 
The labels $(R_c,2I+1,Y_{\phi})$ denote the representation of the new scalar under the  $SU(3)_{c}$, $SU(2)_{L}$ and $U(1)_{Y}$. 
The dashed part of the curves indicates the excluded mass range from the direct collider searches 
for $N_f=1$ (see \ref{app:bounds}). 
For $N_f=N_\text{max}$, these maximal cross sections are given by the dotted curves, 
since we don't know the corresponding experimental bounds (although some of them can be estimated) 
. 
The shaded region is $\sigma_{\gamma\gamma}^\text{Max}>1$fb, which can be consistent with the LHC diphoton excesses. \\

In the top panels, the maximal value of $\sigma_{\gamma\gamma}^{}$ is plotted 
for each diquark model. The models with ${\bf 6^*}$ and {\bf 8} representations under the QCD are not presented 
because these models do not keep the perturbativity up to the Planck scale. 
For diquarks of the {\bf 3} representations, only two representations of $(3,1,-4/3)$ and $(3,3,-1/3)$ with $N_f=1$ 
can be consistent with the perturbativity and the LHC diphoton excess. 
If $N_f>1$, the maximal cross sections are getting larger because these are proportional to $N_f^2$. 
On the other hand, excessively large $N_f$ leads the breakdown of the perturbativity of $\lambda_\phi^{}$ 
before the Planck scale. Even when the perturbativety is kept, 
the allowed maximal value of the scalar trilinear coupling $\mu$ becomes smaller. 
Furthermore, the increase of $N_f$ makes the experimental bound stronger, so that 
the situation does not always become better. 

For the leptoquark extensions (middle panels), some of the models can predict large enough $\sigma_{\gamma\gamma}^{}$, 
however, all the representations suffer from severe direct search constraints. 
For $N_f=1$, no model fulfills the simultaneous requirements of the perturbativity and the LHC diphoton excess 
assuming that the leptoquarks only couple to one generation of the SM fermions. 
The situation is not improved for $N_f>1$, because of the stringent experimental bounds. 
A leptoquark which mainly decays to a light-quark and a tau lepton pair is less constrained 
because of the difficulty of their identifications.  
To our knowledge, no corresponding constraint on such a leptoquark is found, and
it might be consistent with these conditions.

For the dilepton extended models, the values of $\sigma_{\gamma\gamma}^{}$ are relatively small as compared 
with the colored scalar models due to the small degrees of freedom.
%
%
However, thanks to the weaker experimental mass bounds, 
two representations of $(1,3,1)$ and $(1,1,2)$ including the doubly charged Higgs boson 
survive both the experimental and theoretical constraints for $N_f=1$. 
For larger $N_f$, all models can predict a sufficiently large diphoton cross section with lighter dileptons. 
Because of the clean signal of leptonic observable, large $N_f$ would be easily excluded. 
It would be nice if the experimental groups release the results for larger mass regions. 

From these plots, we find that only a few models with $N_f=1$ can explain the LHC diphoton excess 
without conflicting the requirement of the perturbativity up to the Planck scale. 
These conditions typically favor the new scalar field with a higher representation of $SU(2)_L$ 
and a large hypercharge. 
In all the allowed models, the required scalars are relatively light, $M_\phi \lesssim 450$~GeV. 
Its precision measurement would be a good target at the ILC with the CM energy of $1$~TeV. 
The doubly charged scalars in the dilepton multiplets may also be produced at the $e^-e^-$ option 
of the ILC if the relevant Yukawa coupling is sufficiently large. 


\section{Extensions with New Vector-like Fermions}\label{sec:fermion}
In this section, we consider vector-like fermions instead of the scalar multiplets 
discussed in the previous section. 
The new fermions are used to be a source of $g_{X\gamma}^{}$ at one-loop level. 
Similarly to the scalars, 
the representations of the vector-like fermions under the SM gauge group 
are fully determined by allowing the tree level decays to the SM particles. 
Their quantum charges are listed  
in Table~\ref{Tab:VLQs} for vector-like quarks and in Table~\ref{Tab:VLLs} for vector-like leptons.  \\

\begin{table}[tbh]
\centering
\begin{tabular}{c||c|c|c||c|c}
 {}
VLQs & $SU(3)_c$ & $SU(2)_L$ & $U(1)_Y$ & $y_\psi^\text{Max}$ &$ (y_{\psi}^{\text{Max}}N_{f})|_{N_f=N_{\text{Max}}}$\\\hline\hline
$T_0$ & ${\bf 3}$ & ${\bf 1}$ & $2/3$ & $0.84$&$2.5\,|_{N_{\text{Max}^{}}=5}$ \\\hline
$B_0$ & ${\bf 3}$ & ${\bf 1}$ & $-1/3$ & $0.82$&$5.1\,|_{N_{\text{Max}^{}}=13}$\\\hline
$T_1$ & ${\bf 3}$ & ${\bf 3}$ & $2/3$ & $0.80$ &$0.80\,|_{N_\text{Max}^{}=1}$\\\hline
$B_1$ & ${\bf 3}$ & ${\bf 3}$ & $-1/3$ & $0.79$&$0.79\,|_{N_\text{Max}^{}=1}$ \\\hline
$Q_{1/2}$ & ${\bf 3}$ & ${\bf 2}$ & $1/6$ & $0.72$&$1.8|_{N_{\text{Max}^{}}=4}$\\\hline
$T_{1/2}$ & ${\bf 3}$ & ${\bf 2}$ & $7/6$  & LP of $g_{Y}^{}$&LP of $g_{Y}^{}$\\\hline
$B_{1/2}$ & ${\bf 3}$ & ${\bf 2}$ & $-5/6$ & $0.76$&$0.76\,|_{N_\text{Max}^{}=1}$
\end{tabular}
\caption{A list of vector-like quarks which couple to the SM fields. 
For $T_{1/2}$, the Landau pole of $g_{Y}^{}$ appears below the Planck scale even for $N_{f}=1$ \cite{hamada-diphoton}.
}
\label{Tab:VLQs}
\end{table}

\begin{table}[tbh]
\centering
\begin{tabular}{c||c|c|c||c|c}
 {}
VLLs & $SU(3)_c$ & $SU(2)_L$ & $U(1)_Y$ & $y_\psi^\text{Max}$ & $(y_{\psi}^{\text{Max}}N_{f})\,|_{N_f=N_{\text{Max}}}$\\\hline\hline
$N_0$ & ${\bf 1}$ & ${\bf 1}$ & $0$ & $\times$&$\times$\\\hline
$E_0$ & ${\bf 1}$ & ${\bf 1}$ & $-1$& $0.73$&$2.9\,|_{N_{\text{Max}^{}}=7}$\\\hline
$N_1$ & ${\bf 1}$ & ${\bf 3}$ & $0$ & $0.77$&$1.8\,|_{N_{\text{Max}^{}}=3}$ \\\hline
$E_1$ & ${\bf 1}$ & ${\bf 3}$ & $-1$& $0.83$& $1.4\,|_{N_{\text{Max}^{}}=2}$\\\hline
$L_{1/2}$ & ${\bf 1}$ & ${\bf 2}$ & $-1/2$ &$0.68$&$3.6\,|_{N_{\text{Max}^{}}=12}$\\\hline
$E_{1/2}$ & ${\bf 1}$ & ${\bf 2}$ & $-3/2$ &$0.82$&$0.82\,|_{N_f=1}$
\end{tabular}
\caption{A list of vector-like leptons which couple to the SM fields. For $N_{0}$, 
$y_\psi^{\text{Max}}$ is not presented because it is a SM singlet. 
}
\label{Tab:VLLs}
\end{table}

Because all vector-like fermions listed here decay to the Higgs field and the SM fermions, 
they must be color triplet or singlet. 
Therefore, the allowed decay modes are SM fermions plus $W, Z$ or $H_\text{125}$.  
Small Yukawa couplings to the SM fields are sufficient for the prompt decay of the new fermions, 
so that we neglect them 
in the following discussion, i.e., 
masses of each component of vector-like fermions are degenerate 
\footnote{
As a result, the contributions to the electroweak precision test are also negligible.
}. 
The experimental bounds on these vector-like fermion masses are summarized in \ref{app:bounds}. 
For the leptonic extended models, we implicitly assume the existence of a new colored particle 
(with relatively small hypercharge) similarly to the color-singlet scalar extensions. \\

Let us move to the calculation of the diphoton signals of $X$. 
In the following, we denote one of the vector-like fermions in Tables \ref{Tab:VLQs} and \ref{Tab:VLLs} by $\psi$. 
We consider the following two typical Yukawa interactions between $X$ and $\psi$:
\begin{equation}  
{\cal{L}} \ni 
-M_\psi\bar{\psi}\psi 
-\begin{cases}
y_\psi^{}\, X \bar{\psi}\psi&(\text{scalar type})\\
i\,y_\psi^{}\, X \bar{\psi}\gamma_{5}\psi&(\text{pseudoscalar type})
\end{cases}.
\end{equation} 
Because of this new Yukawa coupling, $X$ can decay to $\gamma\gamma$ at the one-loop level. 
The effective coupling $g_{X\gamma}^{}$ is as follows:
\begin{equation}  
g_{X\gamma}^{}
=\frac{2N_{f}\, \alpha \sum_{\psi_i}Q_{\psi_i}^2}{\pi}\frac{y_\psi^{}}{M_{X}}
\times
\begin{cases}f_{S}^{}(M_{\psi}^2/M_X^2) & (\text{scalar type})\label{eq:effcoupling2}\\
f_{PS}^{}(M_{\psi}^2/M_X^2) & (\text{pseudoscalar type})
\end{cases},
\end{equation} 
where
\begin{equation}  
f_{S}^{}(x):=
2 \sqrt{x} \left((1-4 x)\, \arctan\left(\frac{1}{\sqrt{4 x-1}}\right)^2+1\right), 
\end{equation} 
and 
\begin{equation}  \label{fPS}
f_{PS}^{}(x):=
2 \sqrt{x}\, \arctan\left(\frac{1}{\sqrt{4 x-1}}\right)^2.
\end{equation} 
Both functions take the maximal value at the threshold, $f_S^{}(1/4)=1$ and $f_{PS}^{}(1/4)=\pi^2/4$,
and decrease as $f_S^{}\sim1/(3\sqrt{x})$ and $f_{PS}^{}\sim1/(2\sqrt{x})$ for large $x$. 
As a result, for the pseudoscalar case, $\sigma_{\gamma\gamma}$ is about $(\pi^2/4)^2 \approx 6$ times larger than 
that of the scalar case around the threshold, and $(3/2)^2=2.25$ times larger in the heavy fermion limit.

As well as the scalar extension cases, the maximal value of $y_\psi^{}$ can be determined 
by imposing the perturbativity up to the Planck scale. 
The one-loop RGE of $y_\psi^{}$ is given by
\begin{equation}  
\frac{dy_\psi^{}}{dt}=
\frac{y_\psi^{}}{16\pi^2}\left((6 n_\psi^{} N_{f}+3) y_\psi^2 
-6 g_2^2 C_2(n_\psi^{})-6 g_Y^2 Y_\psi^2-8 g_3^2\right)\label{eq:yukawarge1}
\end{equation} 
for vector-like quarks, and 
\begin{equation}  
\frac{dy_\psi^{}}{dt}=
\frac{y_\psi^{}}{16\pi^2}\left((2 n_\psi^{} N_{f}+3) y_\psi^2
-6 g_2^2 C_2(n_\psi^{})-6 g_Y^2 Y_\psi^2\right)\label{eq:yukawarge2}
\end{equation} 
for vector-like leptons. 
Here, $n_\psi^{}(=2I_\psi^{}+1)$ is the $SU(2)_{L}$ representation of $\psi$, 
$C_{2}$ is the Casimir index, and $Y_\psi$ is the $U(1)_{Y}$ hypercharge of $\psi$. 
The RGEs of the gauge couplings are presented in \ref{app:rge}.

In Tables \ref{Tab:VLQs} and \ref{Tab:VLLs}, we list $y_\psi^\text{Max}$ 
(the maximal values of $y_\psi^{}$ at the weak scale) in various vector-like fermion models. 
For these models, $N_{\text{Max}}$ is determined by the perturbativity bound 
of the SM gauge couplings
\footnote{Here, note that $N_{\text{Max}}$ is determined independently of $y_{\psi}$ because the Landau pole of $y_{\psi}$ never appears as long as it is zero at the weak scale.
}. 
We then evaluate $y_\psi^\text{Max}$ for $N_f=N_\text{Max}$. 
The corresponding maximal values of $\sigma_{\gamma\gamma}$ can be calculated 
by substituting $y_\psi^\text{Max}$ in Eq.(\ref{eq:effcoupling2}) and using Eq.(\ref{eq:theoretical cs}). \\

\begin{figure}[h!]
\begin{center}
\begin{tabular}{c}
\begin{minipage}{0.5\hsize}
\begin{center}
\includegraphics[width=8cm]{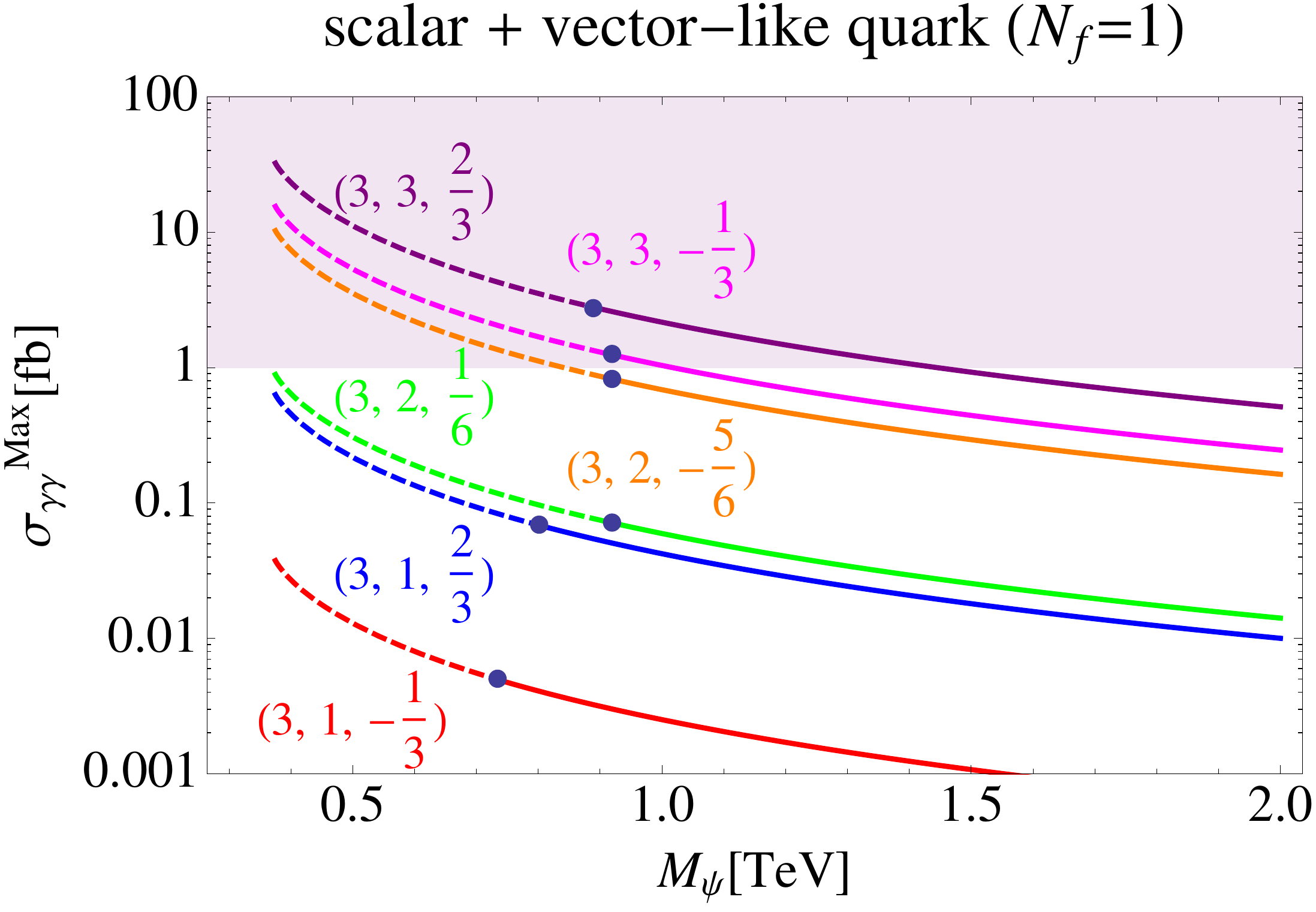}
\end{center}
\end{minipage}
\begin{minipage}{0.5\hsize}
\begin{center}
\includegraphics[width=8cm]{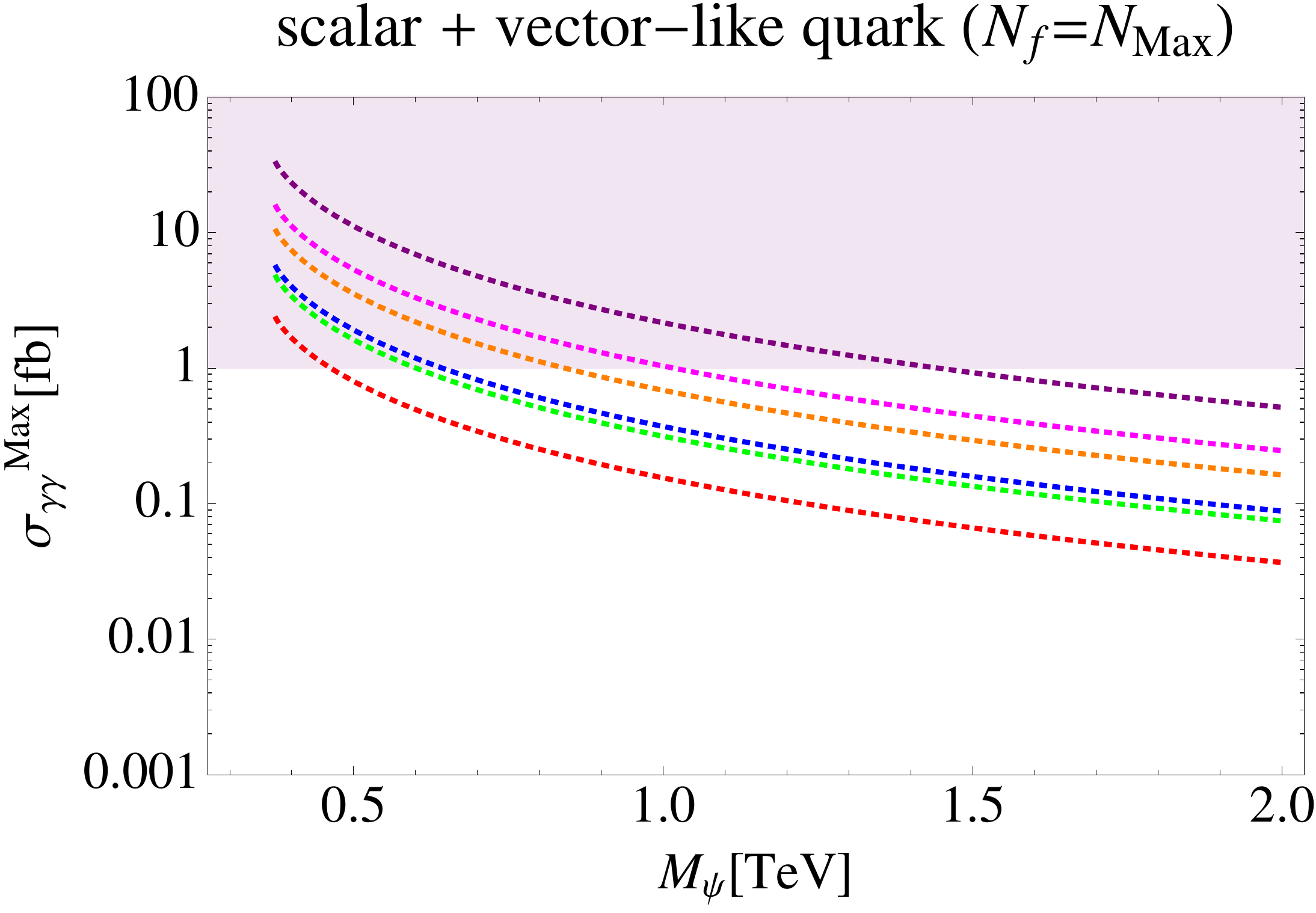}
\end{center}
\end{minipage}
\\
\\
\begin{minipage}{0.5\hsize}
\begin{center}
\includegraphics[width=8cm]{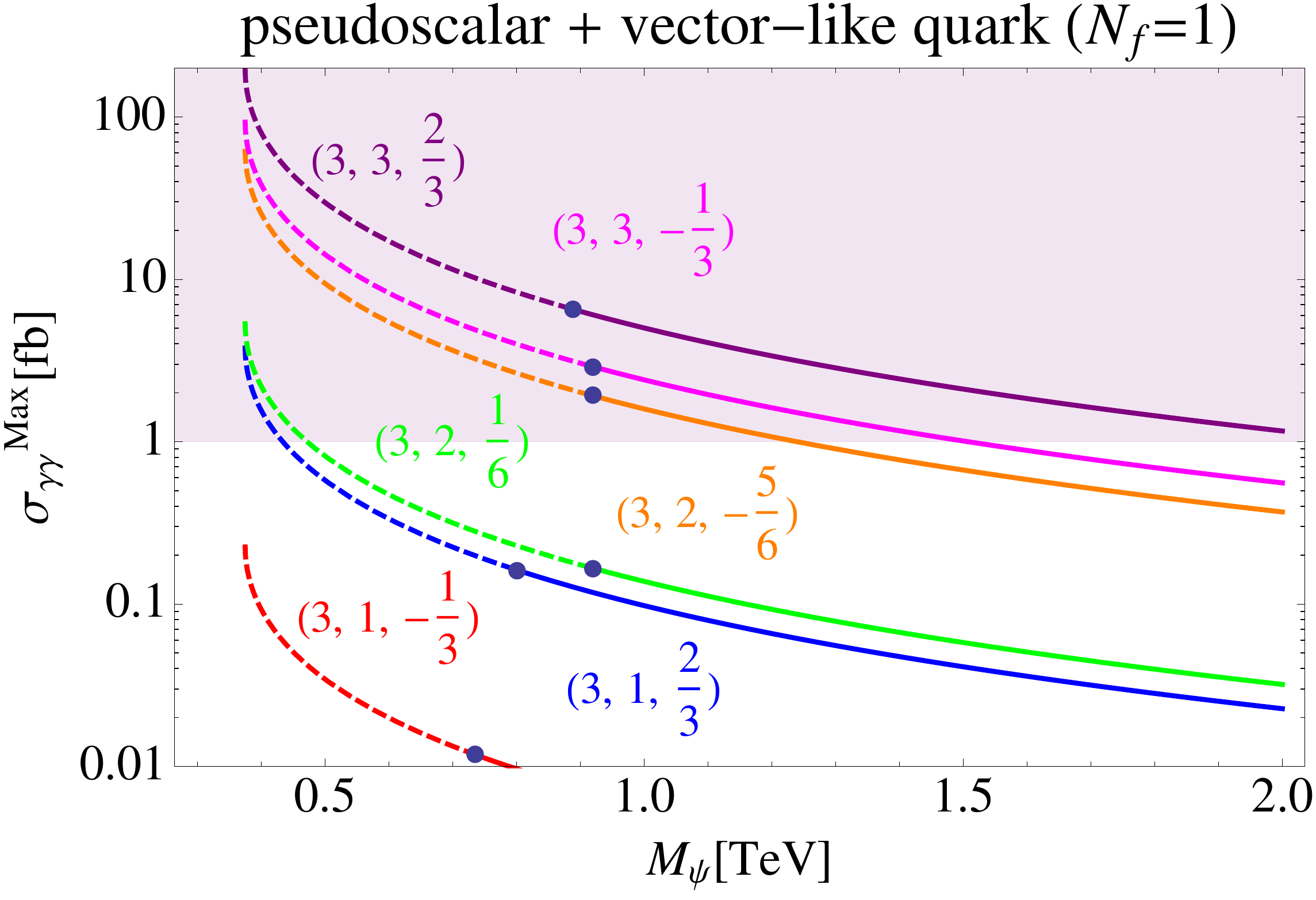}
\end{center}
\end{minipage}
\begin{minipage}{0.5\hsize}
\begin{center}
\includegraphics[width=8cm]{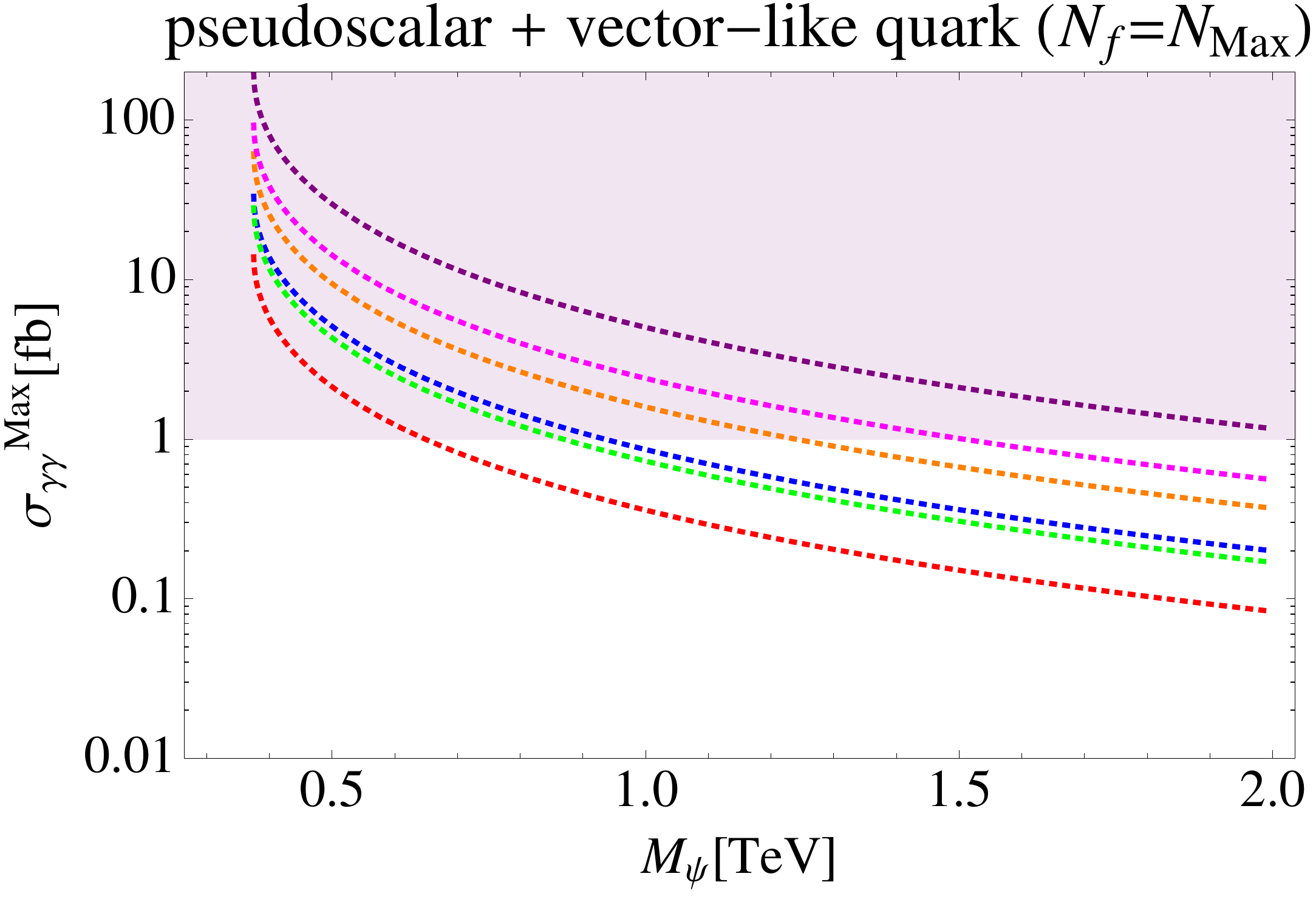}
\end{center}
\end{minipage}
\end{tabular}
\end{center}
\caption{The maximal cross sections $\sigma_{\gamma\gamma}^\text{Max}$ in vector-like quark extension models 
are given as functions of the mass $M_{\psi}$ of the vector-like quarks. 
The curves and shaded region are shown in a similar manner to the scalar extensions. }
\label{fig:quark cs}
\end{figure}

\begin{figure}[h!]
\begin{center}
\begin{tabular}{c}
\begin{minipage}{0.5\hsize}
\begin{center}
\includegraphics[width=8cm]{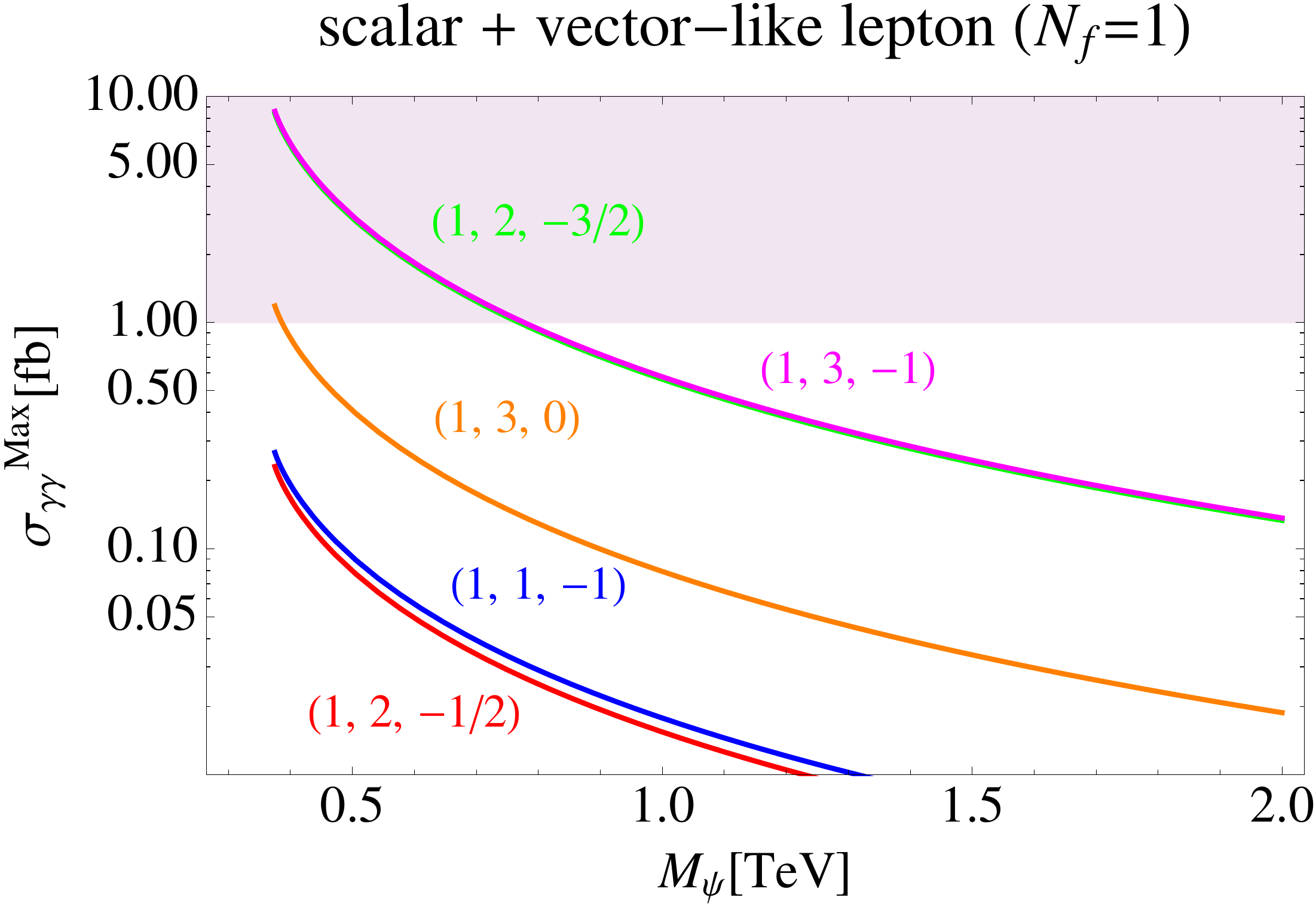}
\end{center}
\end{minipage}
\begin{minipage}{0.5\hsize}
\begin{center}
\includegraphics[width=8cm]{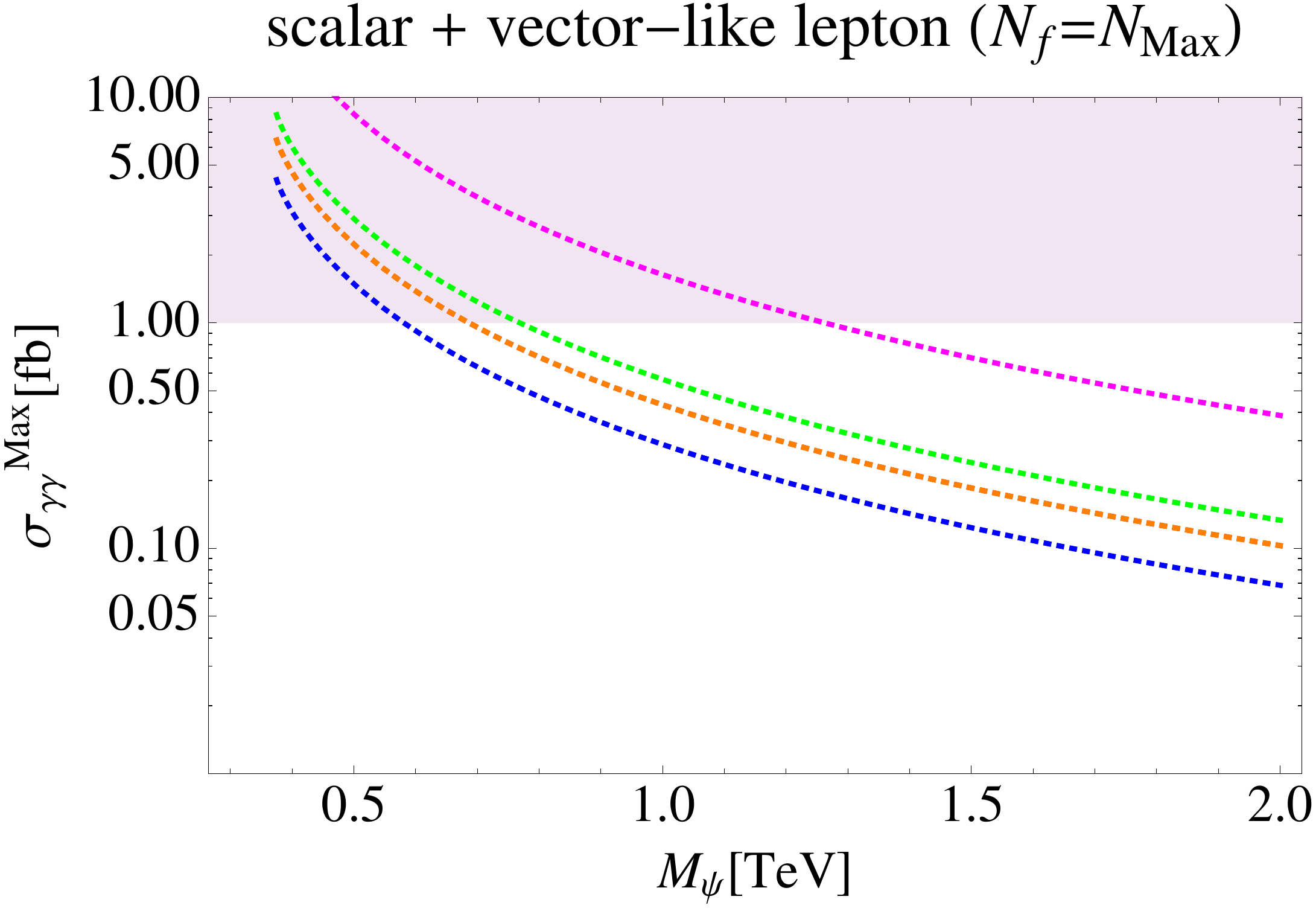}
\end{center}
\end{minipage}
\\
\\
\begin{minipage}{0.5\hsize}
\begin{center}
\includegraphics[width=8cm]{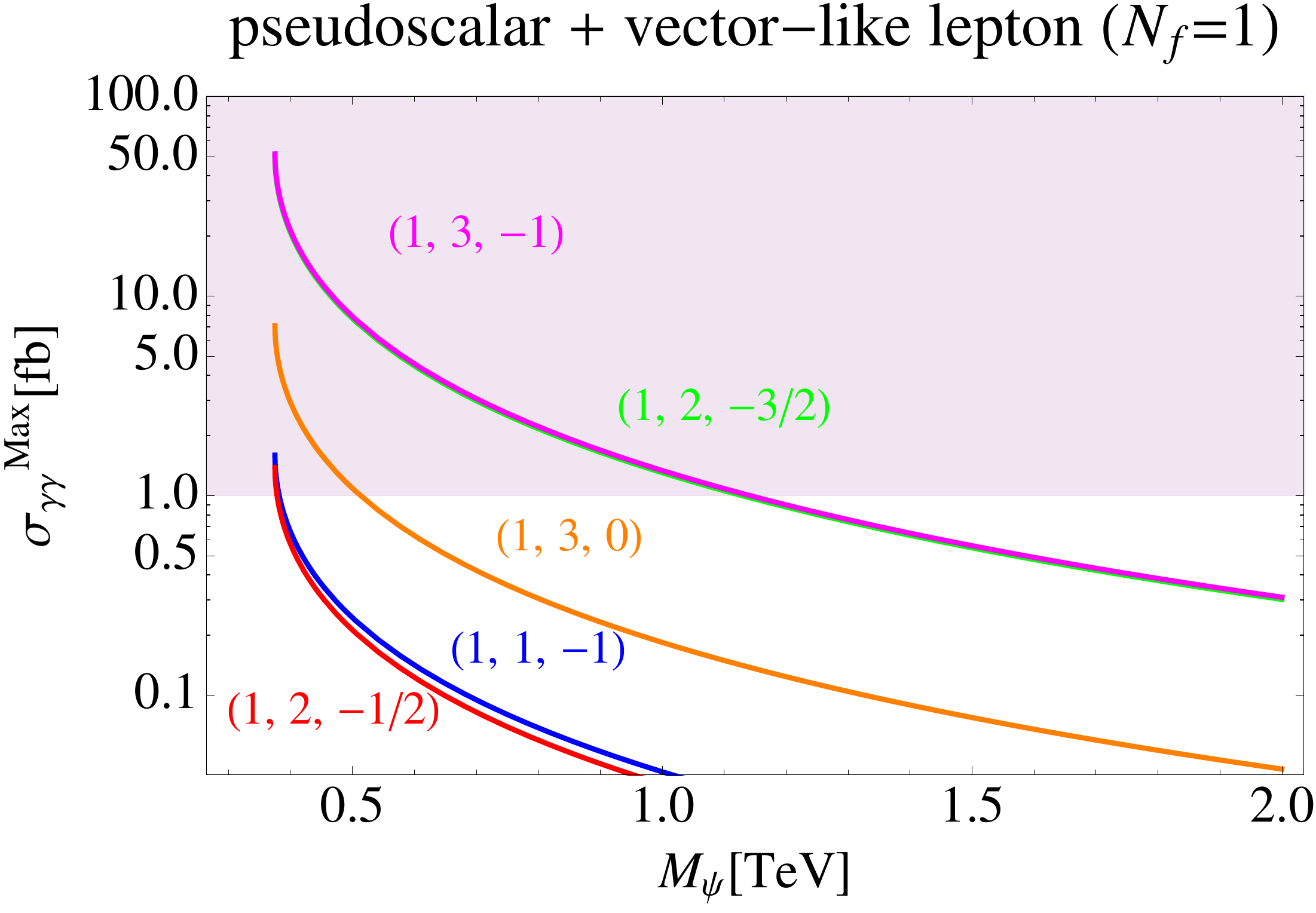}
\end{center}
\end{minipage}
\begin{minipage}{0.5\hsize}
\begin{center}
\includegraphics[width=8cm]{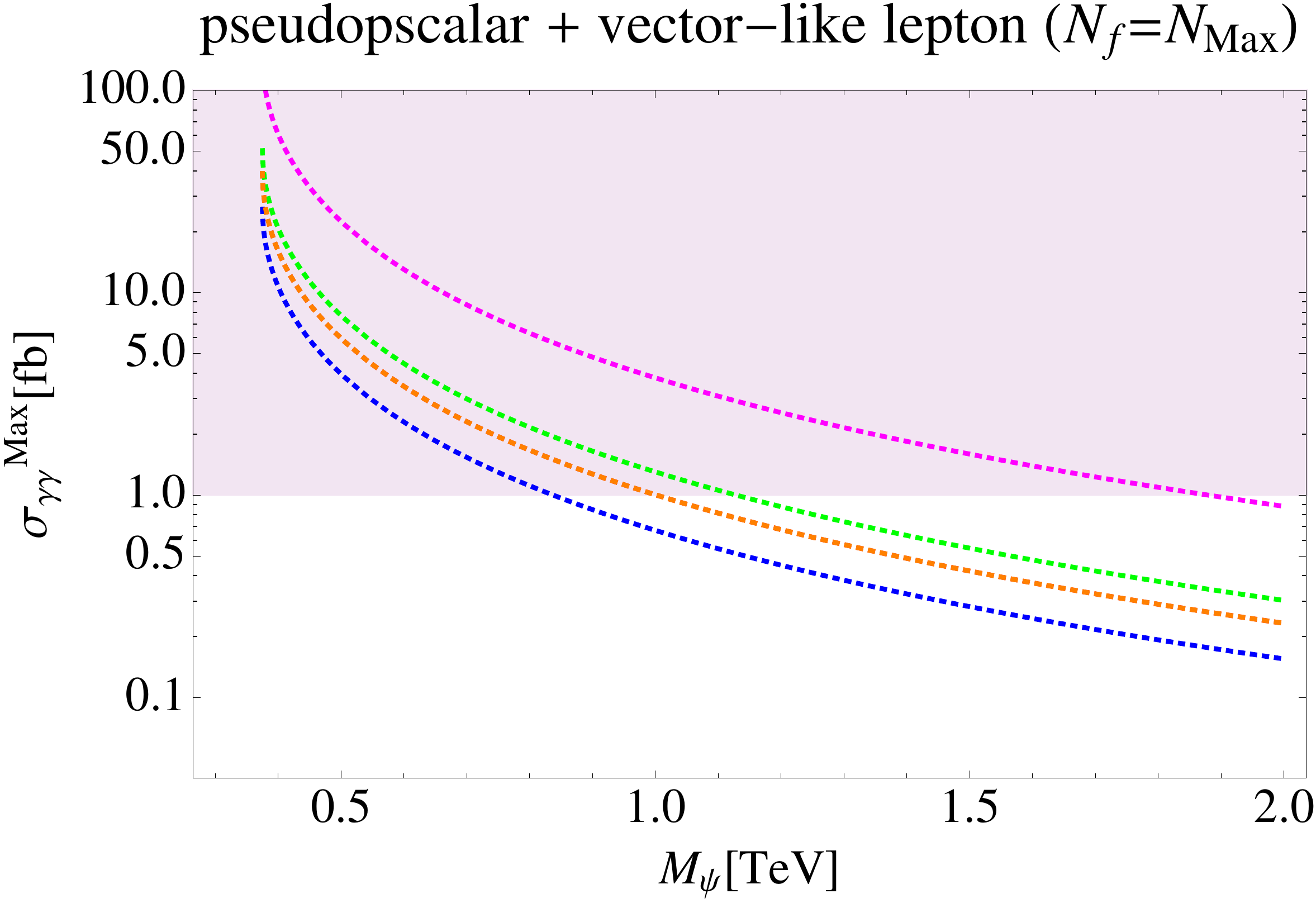}
\end{center}
\end{minipage}
\end{tabular}
\end{center}
\caption{The maximal cross sections $\sigma_{\gamma\gamma}^\text{Max}$ in the vector-like lepton extension models 
are presented as functions of the mass $M_{\psi}$  of the vector-like leptons. }
\label{fig:lepton cs}
\end{figure}

In Figs.~\ref{fig:quark cs} and \ref{fig:lepton cs}, 
we show the maximal cross sections $\sigma_{\gamma\gamma}^\text{Max}$ in the vector-like quark and lepton extension models 
as functions of $M_{\psi}$, respectively. 
For the left (right) panel, the results for $N_f=1\, (N_\text{Max})$ are given similarly to the figures of the scalar extended models. 
The two cases of the scalar and pseudoscalar Yukawa couplings, are presented in the upper and lower panels.

In the vector-like quark extensions, we do not show the result for the $(3,2,7/6)$ representation 
because the Landau pole appears in the running of the hypercharge gauge coupling. 
For the rest, if the Yukawa coupling is the scalar type, two representations of $(3,3,2/3)$ and $(3,3,-1/3)$ are consistent with the
requirements of the perturbativity, the explanation of the LHC diphoton excess, and the LHC direct search bound. 
In the case of the pseudoscalar type, thanks to the enhancement in the loop integral, 
a representation of $(3,2,-5/6)$ is rescued even when $N_f=1$. 
Furthermore, two more representations of $(3,2,1/6)$ and $(3,1,2/3)$ are saved by considering many flavors. 
In contrast to the scalar extensions, the situation gets better for large $N_f$, 
because $y_\psi^\text{Max}$ is not so sensitive to $N_f$.\footnote{The reason is as follows. When $y_{\psi}$ and $N_{f}$ are relatively large, the RGE of $y_{\psi}$ is roughly $dy_{\psi}/dt\sim N_{f}y_{\psi}^3/16\pi^2$ from Eq.(\ref{eq:yukawarge1}) or Eq.(\ref{eq:yukawarge2}). However, by the redefinition $y_{\psi}:=N_{f}^{-1/2}\tilde{y}_{\psi}$, we can see that $\tilde{y}_{\psi}$ follows the RGE when $N_{f}=1$. Therefore, $y_{\psi}^{\text{Max}}$ is roughly given by $N_{f}^{-1/2}\tilde{y}_{\psi}=N_{f}^{-1/2}y_{\psi}^{\text{Max}}|_{N_{f}=1}^{}$, and the effect of $N_{f}$ is not much strong. 
} 
It should be noted that the decoupling behavior is slower than the case of the scalar extension
due to the different loop integral, so that relatively heavy vector-like quark can predict 
sufficiently large $\sigma_{\gamma\gamma}^{}$. 

For vector-like leptons, the experimental mass bounds are very loose because of the small production cross section. 
As we summarize in \ref{app:bounds}, only the electron and muon decay channels have been searched. 
All of the vector-like lepton could be as light as a half of the diphoton resonance mass $M_X$. 
As in the left-top panel of Fig.\ref{fig:lepton cs}, 
the representations of $(1,2,-3/2)$, $(1,3,-1)$ and $(1,3,0)$ can be consistent with all conditions 
for the scalar type Yukawa coupling with $N_f=1$. 
For the pseudoscalar case, all the models survive thanks to the largeness of $f_{PS}$, Eq.~\eqref{fPS}. 
By considering $N_f>1$, the required cross sections are easily satisfied.

These fermionic extensions only demand relatively heavy fermions unlike scalar extended models. 
Such heavy fermions would be beyond the scope at the LHC. 
In order to test the origin of the LHC diphoton excess completely, we may need to go to a further high energy frontier.

\section{Conclusions and Discussion}\label{sec:summary}

We have studied perturbatively safe scenarios for the LHC diphoton excesses. 
The excess can be interpreted by a new scalar boson $X$ 
with $M_X=750$~GeV which decays to a pair of photons. 
A simple scenario is realized by introducing new scalar bosons 
or
fermions charged under the SM gauge group. 
The effective dimension-five interactions of $X$ and photons (gluons) are induced at one-loop of the new particles. 
In order to generate a sufficiently large effective interaction, 
large coupling constants are needed between $X$ and the new particles.
On the other hand, such coupling constants are severely constrained by the theoretical consistencies of the model;
We have investigated the stability bound of the scalar potential 
and the pertubativity conditions up to the Planck scale.

%
We have considered the cases with diquark, leptoquark, and dilepton multiplets for the scalar extensions, 
where all scalars can decay into the SM fermions at tree level. 
Some of the diquark and dilepton extensions can be consistent with all theoretical and experimental requirements  
while all the leptoquark extensions are ruled out due to the stringent direct search bounds. 
Increase of the number of new scalars helps the situation better if we choose appropriate $N_f$.  
%
All allowed models predict the new light scalar boson with a mass less than $450$~GeV, 
which would be a very nice scope of the ILC. 

We have also examined the vector-like fermion extensions, which can mix the SM fermions through the mass matrix. 
For fermionic extensions, most models can simultaneously realize the LHC diphoton excesses and the perturbativity up to the Planck scale without contradicting the direct search mass bounds. 
A larger cross section for $\sigma_{\gamma\gamma}^{}$ can be readily obtained 
by considering the pseudoscalar Yukawa couplings instead of scalar ones.  
The number of new vector-like fermion multiplets is less important to the perturbativity bounds, 
so that many flavors of the new fermions can enhance the diphoton cross section in contrast to the scalar extensions. 
Because of the above two reasons, heavy vector-like fermions are sufficient to explain the LHC diphoton excesses. 

In conclusion, interpretations of the new exciting results by simple scenarios are the first step 
for understanding the beyond the SM. 
As a next step, the requirements of the theoretical consistencies 
such as the stability and the perturbativity are good criteria for constructing a consistent theory.

\section*{Acknowledgement} 
This work is supported by the Grant-in-Aid for Japan Society for the Promotion of Science (JSPS) Fellows 
No.25$\cdot$1107 (YH) and No.27$\cdot$1771 (KK). 
K.T.'s work is supported in part by the MEXT Grant-in-Aid for Scientific Research on Innovative Areas No. 26104704.

\section*{Note added} 
When we have been finishing our work, Ref.~\cite{Bae:2016xni} appeared on the arXiv, which partially overlaps with our work.

\appendix 
\def\thesection{Appendix \Alph{section}}
\section{Lower Mass Bounds for New Particles}\label{app:bounds}

In this Appendix, we summarize the direct search bounds 
on the masses of the new scalar boson and vector-like fermions with $N_f=1$. 
 
\begin{table}[tbh]
\centering
\begin{tabular}{c||c||c|c}
 {}
 & Decay Modes  &
 	$M_{1\text{st},2\text{nd}}$/GeV$\,\gtrsim$ & 
	$M_{3\text{rd}}$/GeV$\,\gtrsim$  \\\hline\hline
$DQ_0^d$ & ${ \begin{matrix} 
	{\overline{u_L^c}}_i{d_L^{}}_j -{\overline{d_L^c}}_i{u_L^{}}_j\\
	\overline{d_R^c}u_R^{} \end{matrix} }$ & 
	${ \begin{matrix} 350_{\bf 3}~(\sim700)_{\bf 6}~\cite{Khachatryan:2014lpa} \\ 350_{\bf 3}~(\sim700)_{\bf 6}~\cite{Khachatryan:2014lpa} \end{matrix} }$ &
	${ \begin{matrix} \text{N.A.} \\ 385_{\bf 3}~(\sim700)_{\bf 6}~\cite{Khachatryan:2014lpa} \end{matrix} }$ \\\hline
$DQ_0^y$ & $\overline{u_R^c}u_R^{}$ & $350_{\bf 3}~(\sim700)_{\bf 6}~\cite{Khachatryan:2014lpa}$ & 
	$350_{\bf 3}~(\sim700)_{\bf 6}~\cite{Khachatryan:2014lpa}$ \\\hline
$DQ_0^u$ & $\overline{d_R^c}d_R^{}$ & $350_{\bf 3}~(\sim700)_{\bf 6}~\cite{Khachatryan:2014lpa}$ & 
	$350_{\bf 3}~(\sim700)_{\bf 6}~\cite{Khachatryan:2014lpa}$\\\hline
${ DQ_1 
	}$ & 
	${ 
	\begin{pmatrix} -\overline{d_L^c}d_L^{} \\ 
		-\overline{u_L^c}d_L^{} -\overline{d_L^c}u_L^{} \\ 
		\overline{u_L^c}u_L^{} \end{pmatrix} }$ &
	${ \begin{pmatrix} 350_{\bf 3}~(\sim700)_{\bf 6}~\cite{Khachatryan:2014lpa} \\ 
		350_{\bf 3}~(\sim700)_{\bf 6}~\cite{Khachatryan:2014lpa} \\ 
		350_{\bf 3}~(\sim700)_{\bf 6}~\cite{Khachatryan:2014lpa} \end{pmatrix} }$ &
	${ \begin{pmatrix} 350_{\bf 3}~(\sim700)_{\bf 6}~\cite{Khachatryan:2014lpa} \\ 
	385_{\bf 3}~(\sim700)_{\bf 6}~\cite{Khachatryan:2014lpa} \\
	350_{\bf 3}~(\sim700)_{\bf 6}~\cite{Khachatryan:2014lpa} \end{pmatrix} }$ \\\hline
${ DQ_{1/2} 
	}$ & 
	${ \begin{matrix} 
	\begin{pmatrix} \overline{u_L^{}}d_R^{} \\ \overline{d_L^{}}d_R^{} \end{pmatrix} \\ 
	\begin{pmatrix} \overline{u_R^{}}d_L^{} \\ -\overline{u_R^{}}u_L^{} \end{pmatrix} \end{matrix} }$ &
	${ \begin{matrix} 
	\begin{pmatrix} (\sim700)_{\bf 8}~\cite{Khachatryan:2014lpa} \\ 
	(\sim700)_{\bf 8}~\cite{Khachatryan:2014lpa} \end{pmatrix} \\ 
	\begin{pmatrix} (\sim700)_{\bf 8}~\cite{Khachatryan:2014lpa} \\ 
	(\sim700)_{\bf 8}~\cite{Khachatryan:2014lpa} \end{pmatrix} \end{matrix} }$ &  
	${ \begin{matrix} 
	\begin{pmatrix} (\sim700)_{\bf 8}~\cite{Khachatryan:2014lpa} \\ 
	(\sim700)_{\bf 8}~\cite{Khachatryan:2014lpa} \end{pmatrix} \\ 
	\begin{pmatrix} (\sim700)_{\bf 8}~\cite{Khachatryan:2014lpa} \\ 
	650_{\bf 8}~\cite{Calvet:2012rk}~(\sim700)_{\bf 8}~\cite{Khachatryan:2014lpa} \end{pmatrix} \end{matrix} }$
\end{tabular}
\caption{Lower mass bounds for the diquark multiplets. 
The numbers in the parenthesis are our estimated bounds from the scaling of the cross sections for the color triplet diquark.  
}
\label{Tab:MB-DQs}
\end{table}

%
In Table~\ref{Tab:MB-DQs}, collider search bounds for the scalar diquarks are listed. 
The quantum charges of diquarks in the first column are defined in Table~\ref{Tab:DQs}. 
The second column presents possible decay modes to the SM particles. 
The experimental mass bounds strongly depend on the structure of the Yukawa matrices. 
However, there is no way to specify the structure of the Yukawa matrices without additional assumptions, 
so that we focus on two classes of the diquarks : 
1) diquarks which only decay to the first two generation quarks, 
2) diquarks which only decay to the third generation quarks. 
We here do not consider mixture scenarios for simplicity. 
Note that if the diquark decays into the third generation quarks, 
bottom- or top-tagging methods can be applied in order to enhance signal efficiencies. 

The scalar diquarks can be produced from the $qq$ fusion if the Yukawa coupling is sufficiently large. 
Since the bounds are negligible for the smaller Yukawa coupling, we here only take into account the bounds 
from pair production through the QCD interaction. 
The lower mass bound for the color triplet diquarks can be estimated 
from the squark pair production process in the supersymmetric standard model 
with the R-parity violation~\cite{Khachatryan:2014lpa}. 
With (without) the b-tagging, the lower mass bounds is $385(350)$~GeV. 
For color sextet (octet) diquarks, the production cross section is roughly 
12.5 (13.5) times larger than that for color triplets. 
Then, we can read the lower mass bounds for the color sextet and the octet to be about $700$~GeV. 
Dedicated searches for the color octet scalar are found in Refs.~\cite{Calvet:2012rk} and \cite{ATLAS:2012ds}, 
which give a weaker bound, $650$~GeV, as compared to the scaled ones. 

\begin{table}[tbh]
\centering
\begin{tabular}{c||c||c|c|c}
 {}
 & Decay Modes &  
 	$M_{1\text{st}}$/GeV$\,\gtrsim$ & 
	$M_{2\text{nd}}$/GeV$\,\gtrsim$ & 
	$M_{3\text{rd}}$/GeV$\,\gtrsim$  \\\hline\hline
$S_0^{d*}$ &
	${ \begin{matrix}
			\overline{u_L^c}e_L^{} -\overline{d_L^c}\nu_L^{} \\
		\overline{u_R^c}e_R^{} \end{matrix} }$ & 
	${ \begin{matrix} 850~\cite{Khachatryan:2015vaa} \\ 1100~\cite{Khachatryan:2015vaa} \end{matrix} }$ &
	${ \begin{matrix} 760~\cite{Khachatryan:2015vaa} \\ 1080~\cite{Khachatryan:2015vaa} \end{matrix} }$ &
	${ \begin{matrix} 560~\cite{Khachatryan:2015bsa} \\ 685~\cite{Khachatryan:2015bsa} \end{matrix} }$
	\\\hline
$S_0^{y*}$ & 
	$\overline{d_R^c}e_R^{}$ & 1100~\cite{Khachatryan:2015vaa} & 1080~\cite{Khachatryan:2015vaa} & 740~\cite{Khachatryan:2014ura} \\\hline
${ S_1^* 
	}$ & 
	${ 
		\begin{pmatrix} 
		-\overline{d_L^c}e_L^{} \\ 
		-\overline{u_L^c}e_L^{}-\overline{d_L^c}\nu_L^{} \\
		\overline{u_L^c}\nu_L^{} \end{pmatrix} }$ & 
	${ \begin{pmatrix} 1100~\cite{Khachatryan:2015vaa} \\ 850~\cite{Khachatryan:2015vaa} \\ 440~\cite{Aad:2014wea} \end{pmatrix} }$ &
	${ \begin{pmatrix} 1080~\cite{Khachatryan:2015vaa} \\ 760~\cite{Khachatryan:2015vaa} \\ 540~\cite{Aad:2015gna} \end{pmatrix} }$ &
	${ \begin{pmatrix} 740~\cite{Khachatryan:2014ura} \\ 560~\cite{Khachatryan:2015bsa} \\ 750~\cite{CMS:2015kza} \end{pmatrix} }$ \\\hline
${ S_{1/2} 
	}$ & 
	${ \begin{matrix} 
		\begin{pmatrix} \overline{u_L^{}}e_R^{} \\ \overline{d_L^{}}e_R^{} \end{pmatrix} \\ 
		\begin{pmatrix} \overline{u_R^{}}e_L^{} \\ -\overline{u_R^{}}\nu_L^{} \end{pmatrix} \end{matrix} }$ & 
	${ \begin{matrix} \begin{pmatrix} 1100~\cite{Khachatryan:2015vaa} \\ 1100~\cite{Khachatryan:2015vaa} \end{pmatrix} \\ 
	\begin{pmatrix} 1100~\cite{Khachatryan:2015vaa} \\ 440~\cite{Aad:2014wea} \end{pmatrix} \end{matrix} }$ &
	${ \begin{matrix} \begin{pmatrix} 1080~\cite{Khachatryan:2015vaa} \\ 1080~\cite{Khachatryan:2015vaa} \end{pmatrix} \\ 
	\begin{pmatrix} 1080~\cite{Khachatryan:2015vaa} \\ 540~\cite{Aad:2015gna} \end{pmatrix} \end{matrix} }$ &
	${ \begin{matrix} \begin{pmatrix} 685~\cite{Khachatryan:2015bsa} \\ 740~\cite{Khachatryan:2014ura} \end{pmatrix} \\ 
	\begin{pmatrix} 685~\cite{Khachatryan:2015bsa} \\ 750~\cite{CMS:2015kza} \end{pmatrix} \end{matrix} }$ \\\hline
${ S^q_{1/2} 
	}$ & 
	${ 
		\begin{pmatrix} \overline{d_R^{}} e_L^{} \\ -\overline{d_R^{}} \nu_L^{} \end{pmatrix} }$ & 
	${ \begin{pmatrix} 1100~\cite{Khachatryan:2015vaa} \\ 440~\cite{Aad:2014wea} \end{pmatrix} }$ &
	${ \begin{pmatrix} 1080~\cite{Khachatryan:2015vaa} \\ 440~\cite{Aad:2014wea} \end{pmatrix} }$ &
	${ \begin{pmatrix} 740~\cite{Khachatryan:2014ura} \\ 700~\cite{CMS:2014nia} \end{pmatrix} }$ \\ \hline 
$R_0^{u*}$ & 
	$\overline{u_R^c}\nu_R^{}$ & 440~\cite{Aad:2014wea} & 540~\cite{Aad:2015gna} & 750~\cite{CMS:2015kza} \\\hline
$R_0^{d*}$ & 
	$\overline{d_R^c}\nu_R^{}$ & 440~\cite{Aad:2014wea} & 440~\cite{Aad:2014wea} & 700~\cite{CMS:2014nia} \\\hline
${ R_{1/2}^* 
	}$ & 
	${ 
		\begin{pmatrix} \overline{\nu_R^{}} d_L^{} \\ -\overline{\nu_R^{}} u_L^{} \end{pmatrix} }$ & 
	${ \begin{pmatrix} 440~\cite{Aad:2014wea} \\ 440~\cite{Aad:2014wea} \end{pmatrix} }$ &
	${ \begin{pmatrix} 440~\cite{Aad:2014wea} \\ 540~\cite{Aad:2015gna} \end{pmatrix} }$ &
	${ \begin{pmatrix} 700~\cite{CMS:2014nia} \\ 750~\cite{CMS:2015kza} \end{pmatrix} }$
\end{tabular}
\caption{Lower mass bounds for the leptoquark multiplets. }
\label{Tab:MB-LQs}
\end{table}

%
In Table~\ref{Tab:MB-LQs}, we collect the direct search bounds for the scalar leptoquarks. 
The charge assignments are given in Table~\ref{Tab:LQs}. 
Flavor changing Yukawa interactions of the leptoquarks are severely constrained by 
non-observation of lepton flavor violation (LFV) in low energy data~\cite{Carpentier:2010ue}.  
Therefore, we classify the leptoquarks by the SM fermion generation\footnote{
A leptoquark which couples to one lepton flavor and one quark flavor can avoid 
stringent bounds from the LFV. 
}. 
Apart from LFV constraints, 
masses of leptoquarks are also strongly constrained by direct searches at the LHC, 
because of large enough production cross sections by strong interaction and clean leptonic signatures. 
Even if the decay products contain neutrinos, missing energies play an important role as a trigger. 

Bounds for the first two generation leptoquarks, which have a decay mode containing a charged lepton, 
are presented in Ref.~\cite{Khachatryan:2015vaa,Aad:2015caa}. 
These constraints come from the combination of $\ell\ell jj$ and $\ell\nu jj$ final states 
in the pair production of a color triplet scalar.  
%
There is another constraint from a single production through the Yukawa coupling~\cite{Khachatryan:2015qda}, 
which can be easily evaded by assuming small Yukawa couplings. 
For the third generation leptoquarks, 
similar bounds are found in Ref.~\cite{Khachatryan:2014ura,Chatrchyan:2013xna,Khachatryan:2015bsa}. 
Although, tau leptons are difficult to identify in the detectors due to the missing neutrinos in their decay chains, 
further selection cuts such as $b$- or $t$- tagging make the bounds stronger. 
The bounds for the leptoquark related to $\nu_R^{}$ can be estimated from the squark pair production 
in the massless neutralino limit. 
For the first two generation, $M_\phi > 440~(540)$~GeV is obtained\cite{Aad:2014wea,Chatrchyan:2014lfa}
(the charm-tagging technique is applied in Ref.~\cite{Aad:2015gna}). 
For the third generation associated with the bottom (top) quark, 
$M_\phi > 700~(750)$~GeV is provided in Ref.~\cite{CMS:2014nia} (\cite{CMS:2015kza}).

%

\begin{table}[tbh]
\centering
\begin{tabular}{c||c||c|c|c}
 {}
 & Decay Modes &
 	$M_{1\text{st}}$/GeV$\gtrsim$ & 
	$M_{2\text{nd}}$/GeV$\gtrsim$ & 
	$M_{3\text{rd}}$/GeV$\gtrsim$   \\\hline\hline
$h_0^+$ & 
	${ \begin{matrix}
			\overline{\nu_L^c}_i{e_L^{}}_j -\overline{e_L^c}_i{\nu_L^{}}_j \\
		\overline{e_R^c}\nu_R^{} \end{matrix} }$ & 
	${ \begin{pmatrix} \text{N.A.} \\ 200~\cite{Aad:2014vma} \end{pmatrix} }$ &
	${ \begin{pmatrix} \text{N.A.} \\ 210~\cite{Aad:2014vma} \end{pmatrix} }$ &
	${ \begin{pmatrix} \text{N.A.} \\ 90.6~\cite{Aad:2014yka} \end{pmatrix} }$
	\\\hline
$h_0^{++}$ & $\overline{e_R^c}e_R^{}$ & 374~\cite{ATLAS:2014kca} & 438~\cite{ATLAS:2014kca} & 169~\cite{Chatrchyan:2012ya} \\\hline
${ \Delta_1 =
	}$ & 
	${ 
		\begin{pmatrix} 
		-\overline{e_L^c}e_L^{} \\ 
		-\overline{\nu_L^c}e_L^{}-\overline{e_L^c}\nu_L^{} \\
		\overline{\nu_L^c}\nu_L^{} \end{pmatrix} }$ & 
	${ \begin{pmatrix} (551)~\cite{ATLAS:2014kca} \\ (200)~\cite{Aad:2014vma} \\ \sim M_Z/2 \end{pmatrix} }$ &
	${ \begin{pmatrix} (516)~\cite{ATLAS:2014kca} \\ (210)~\cite{Aad:2014vma} \\ \sim M_Z/2 \end{pmatrix} }$ &
	${ \begin{pmatrix} (204)~\cite{Chatrchyan:2012ya} \\ (90.6)~\cite{Aad:2014yka} \\ \sim M_Z/2 \end{pmatrix} }$ \\\hline
${ \Phi_{1/2} 
	}$ & 
	${ \begin{matrix} 
		\begin{pmatrix} \overline{\nu_L^{}}e_R^{} \\ \overline{e_L^{}}e_R^{} \end{pmatrix} \\ 
		\begin{pmatrix} \overline{\nu_R^{}}e_L^{} \\ -\overline{\nu_R^{}}\nu_L^{} \end{pmatrix} \end{matrix} }$ & 
	${ \begin{matrix} \begin{pmatrix} (270)~\cite{Aad:2014vma} \\ \sim M_Z/2 \end{pmatrix} \\ 
	\begin{pmatrix} (270)~\cite{Aad:2014vma} \\ \sim M_Z/2 \end{pmatrix} \end{matrix} }$ &
	${ \begin{matrix} \begin{pmatrix} (270)~\cite{Aad:2014vma} \\ \sim M_Z/2 \end{pmatrix} \\ 
	\begin{pmatrix} (270)~\cite{Aad:2014vma} \\ \sim M_Z/2 \end{pmatrix} \end{matrix} }$ &
	${ \begin{matrix} \begin{pmatrix} (94)~\cite{Abbiendi:2013hk} \\ \sim M_Z/2 \end{pmatrix} \\ 
	\begin{pmatrix} (94)~\cite{Abbiendi:2013hk} \\ \sim M_Z/2 \end{pmatrix} \end{matrix} }$ \\\hline
$s_0^{}$ & $\overline{\nu_R^c}\nu_R^{}$ & --- & --- & --- 
\end{tabular}
\caption{Lower mass bounds for the dilepton multiplets. 
For the singly charged singlet, more than 2 flavors are required in order to have the Yukawa couplings. 
The numbers in the parenthesis indicate the existence of other decay modes, which can weaken the lower mass bounds. 
}
\label{Tab:MB-DLs}
\end{table}

%
In Table~\ref{Tab:MB-DLs}, direct search bounds for the scalar dileptons are summarized. 
The representations under the SM gauge group are given in Table~\ref{Tab:DLs}. 
In the second column, we list the possible Yukawa interaction with the SM leptons and $\nu_R^{}$. 
The LHC bounds for scalar dileptons are generally loose  
since its production mechanism relies on the weak interaction. 
Unlike colored scalar multiplets, dileptons can decay to a pair of SM bosons, $W/Z$ and $h_{125}^{}$. 
Such decay modes can weaken the direct search bounds on dilepton masses. 
Furthermore, the mass bounds for an extra doublet $\Phi$ strongly depend on the other Yukawa couplings. 
The numbers in the parenthesis indicate the presence of such additional decay modes, 
which may weaken the lower mass bounds. 
Several dilepton fields are sometimes discussed in the models for small neutrino masses. 
In such a case, the structure of the Yukawa interactions are constrained by the model. 
We here classify the dileptons by the SM lepton generation for simplicity.

A bound for singly charged scalar bosons is read from the slepton mass bounds 
in massless neutralino limit~\cite{Aad:2014vma,Khachatryan:2014qwa,Aad:2014yka}. 
Note that the gauge couplings of singly charged Higgs bosons in $h_0^\pm$ and $\Delta_1$ are 
the same as those of the right-handed sleptons. 
The singly charged scalar bosons in the doublet have the same charges with the left-handed slepton. 
In the Type-II two Higgs doublet model assuming ${\mathcal B}(\tau\nu+cs)=1$,  
a weaker bound for a singly charged scalar is given by LEP~\cite{Abbiendi:2013hk}. 
The bounds for the neutral component in the doublet have large ambiguity 
due to the possible Yukawa coupling to the quarks. 
We have, at least, a bound of $\sim M_Z/2$ from the precisely measured $Z$ boson decay width.
Potentially strong bounds are found for doubly charged scalar bosons by assuming 
the same-sign two lepton decay modes\cite{ATLAS:2014kca,Aad:2014hja}. 
If the doubly charged scalars decay to a pair of $W$ boson (it is possible for the triplet dilepton case), 
the lower mass bound is only $84$ GeV~\cite{Kanemura:2014ipa} from the inclusive search mode. 
No bound is obtained for the SM singlet scalar $s_0^{}$.

\begin{table}[tbh]
\centering
\begin{tabular}{c||c||c}
 {}
 & Decay Modes & 
 	$M_{3\text{rd}}$/GeV$\gtrsim$\\\hline\hline
$T_0$ & $ 
	\overline{t_L^{}}\frac{H+i\,G^0}{\sqrt2}+\overline{b_L^{}}i\,G^-$ &
	800~\cite{Aad:2015kqa} \\\hline
$B_0$ & $ 
	\overline{t_L^{}} i\,G^+ +\overline{b_L^{}} \frac{H-i\, G^0}{\sqrt2}$ & 
	735~\cite{Aad:2015kqa} \\\hline
${ T_1 
	}$ & 
	${ 
		\begin{pmatrix} 
		\overline{t_L^{}} i\,G^- \\ 
		\overline{t_L^{}} \frac{H+i\,G^0}{\sqrt2} +\overline{b_L^{}} i\,G^- \\
		\overline{b_L^{}} \frac{H+i\,G^0}{\sqrt2} \end{pmatrix} }$ &
		${ \begin{pmatrix} 890~\cite{Khachatryan:2015gza} \\ 800~\cite{Aad:2015kqa} \\ 755~\cite{Aad:2014efa} \end{pmatrix} }$
		\\\hline
${ B_1 
	}$ & 
	${ 
		\begin{pmatrix} 
		\overline{t_L^{}} \frac{H-i\,G^0}{\sqrt2} \\ 
		\overline{t_L^{}} i\,G^+ +\overline{b_L^{}} \frac{H-i\,G^0}{\sqrt2} \\
		\overline{b_L^{}} i\,G^+ \end{pmatrix} }$ &
		${ \begin{pmatrix} 855~\cite{Aad:2015kqa} \\ 735~\cite{Aad:2015kqa} \\ 920~\cite{Khachatryan:2015oba} \end{pmatrix} }$ \\\hline
${ Q_{1/2} 
	}$ & 
	${ \begin{matrix} 
		\begin{pmatrix} \overline{t_R^{}} \frac{H-i\, G^0}{\sqrt2} \\ -\overline{t_R^{}}  i\,G^+  \end{pmatrix} \\ 
		\begin{pmatrix} \overline{b_R^{}} i\,G^- \\ -\overline{b_R^{}} \frac{H+i\, G^0}{\sqrt2} \end{pmatrix} \end{matrix} }$ & 
	${ \begin{matrix} 
	\begin{pmatrix} 855~\cite{Aad:2015kqa} \\ 890~\cite{Khachatryan:2015gza} \end{pmatrix} \\ 
	\begin{pmatrix} 920~\cite{Khachatryan:2015oba} \\ 755~\cite{Aad:2014efa} \end{pmatrix} \end{matrix} }$ \\\hline
${ T_{1/2} 
	}$ & 
	${ 
	\begin{pmatrix} \overline{t_R^{}} i\,G^- \\ -\overline{t_R^{}} \frac{H+i\, G^0}{\sqrt2} \end{pmatrix} }$ &
	${ \begin{pmatrix} 890~\cite{Khachatryan:2015gza} \\ 855~\cite{Aad:2015kqa} \end{pmatrix} }$ \\\hline
${ B_{1/2} 
	}$ & 
	${ 
	\begin{pmatrix} \overline{b_R^{}} \frac{H-i\, G^0}{\sqrt2} \\ -\overline{b_R^{}} i\,G^+ \end{pmatrix} }$ &
	${ \begin{pmatrix} 755~\cite{Aad:2014efa} \\ 920~\cite{Khachatryan:2015oba} \end{pmatrix} }$ 
\end{tabular}
\caption{Lower mass bounds for the vector-like quark multiplets.}
\label{Tab:MB-VLQs}
\end{table}

%
In Table~\ref{Tab:MB-VLQs}, lower mass bounds for the third generation vector-like quarks are listed. 
The quantum charges of them are defined in Table~\ref{Tab:VLQs}. 
The second column is a guideline for estimating the decay branching ratios of vector-like fermions, 
where the Higgs field is expanded as $\Phi=(i\,G^+, \frac{H-i\,G^0}{\sqrt2})^T$. 
The NG boson fields $G^0, G^\pm$ indicate the decays to $Z, W$ bosons in the heavy mass limit 
through the electroweak equivalence theorem. 
 
Thanks to the bottom- and/or top-tagging technique,  stronger lower bounds of 
$735$--$855$~GeV are obtained for the vector-like bottom ($Q_\psi=-1/3$) and top quarks ($Q_\psi=2/3$) 
depending on the branching fractions~\cite{Aad:2015kqa,Aad:2014efa}. 
Slightly stronger bounds of $890$--$920$~GeV
are given for the vector-like fermions with the charge $5/3$ and $-4/3$, 
since the flipped charged $W$ boson is a good discriminant from the background events 
as compared with the conventional vector-like top and bottom quark decays to 
the $W$ boson~\cite{Khachatryan:2015gza,Khachatryan:2015oba}.  
If a vector-like quark decays into the a light quark (the first two generations), 
the bounds become weaker. 
For example, a vector-like quark of ${\mathcal B}(Wq)=1\,(0.5)$ is excluded 
from $320\,(390)$~GeV to $690\,(410)$~GeV~\cite{Aad:2015tba}.

{
\renewcommand\thefootnote{\alph{footnote}}
\begin{table}[tbh]
\centering
\begin{tabular}{c||c||c|c}
 & Decay Modes & 
	$M_{1\text{st}}$/GeV$\gtrsim$ & $M_{2\text{st}}$/GeV$\gtrsim$ \\\hline\hline
$N_0$ & $
	\overline{\nu_L^{}}\frac{H+i\,G^0}{\sqrt2} +\overline{e_L^{}}\, i\,G^-$ & 
	--- & --- \\\hline
$E_0$ & $
	\overline{\nu_L^{}}\, i\,G^+ +\overline{e_L^{}}\frac{H-i\,G^0}{\sqrt2}$ &
	176\footnotemark[1]~\cite{Aad:2015dha} & 
	168\footnotemark[2]~\cite{Aad:2015dha} \\\hline
${ N_1 
	}$ & 
	${ 
	\begin{pmatrix} \overline{\nu_L^{}}\, i\, G^- \\ 
	\overline{\nu_L^{}}\frac{H+i\,G^0}{\sqrt2} +\overline{e_L^{}}\, i\,G^- \\ 
	\overline{e_L^{}}\frac{H+i\,G^0}{\sqrt2} \end{pmatrix} }$ &
	430~\cite{Aad:2015dha} & 
	468\footnotemark[3]~\cite{Aad:2015dha} \\\hline
${ E_1
	}$ & 
	${ 
	\begin{pmatrix} \overline{\nu_L^{}}\frac{H-i\,G^0}{\sqrt2} \\ 
	\overline{\nu_L^{}}\, i\,G^+ +\overline{e_L^{}}\frac{H-i\,G^0}{\sqrt2} \\ 
	\overline{e_L^{}}\, i\,G^+ \end{pmatrix} }$ &
	${ \begin{pmatrix} \sim M_Z/2 \\ 176\footnotemark[1]~\cite{Aad:2015dha} \\ \sim M_Z/2 \end{pmatrix} }$ &
	${ \begin{pmatrix} \sim M_Z/2 \\ 168\footnotemark[2]~\cite{Aad:2015dha} \\ \sim M_Z/2 \end{pmatrix} }$ \\\hline
${ L_{1/2} 
	}$ & 
	${ \begin{matrix} 
		\begin{pmatrix} \overline{\nu_R^{}} \frac{H-i\,G^0}{\sqrt2} \\ 
			-\overline{\nu_R^{}}\, i\, G^+ \end{pmatrix} \\ 
		\begin{pmatrix} \overline{\nu_R^c} \frac{H-i\,G^0}{\sqrt2} \\ 
			-\overline{\nu_R^c}\, i\, G^+ \end{pmatrix}\\
		\begin{pmatrix} \overline{e_R^{}}\, i\, G^- \\ 
			-\overline{e_R^{}} \frac{H+i\,G^0}{\sqrt2} \end{pmatrix}
		\end{matrix}}$ & 
		${ \begin{matrix} 
		\begin{pmatrix} 300~\cite{Khachatryan:2014mma} \\ 101.2~\cite{Achard:2001qw} \end{pmatrix} \\
		\begin{pmatrix} 300~\cite{Khachatryan:2014mma} \\ 101.2~\cite{Achard:2001qw} \end{pmatrix} \\
		\begin{pmatrix} 102.6~\cite{Achard:2001qw} \\ \sim M_Z/2 \end{pmatrix} 
		\end{matrix}}$ & 
		${ \begin{matrix} 
		\begin{pmatrix} 300~\cite{Khachatryan:2014mma} \\ 101.2~\cite{Achard:2001qw} \end{pmatrix} \\
		\begin{pmatrix} 300~\cite{Khachatryan:2014mma} \\ 101.2~\cite{Achard:2001qw} \end{pmatrix} \\
		\begin{pmatrix} 102.7~\cite{Achard:2001qw} \\ \sim M_Z/2 \end{pmatrix} 
		\end{matrix}}$ \\\hline
${ E_{1/2} 
	}$ & 
	${ 
	\begin{pmatrix} \overline{e_R^{}} \frac{H-i\,G^0}{\sqrt2} \\ 
	-\overline{e_R^{}}\, i\, G^+ \end{pmatrix} }$ &
	${ \begin{pmatrix}  \sim M_Z/2 \\ \sim M_Z/2 \end{pmatrix} }$ &
	${ \begin{pmatrix}  \sim M_Z/2 \\ \sim M_Z/2 \end{pmatrix} }$
\end{tabular}
\caption{Lower mass bounds for the vector-like lepton multiplets. 
}
\label{Tab:MB-VLLs}
\end{table}


%
In Table~\ref{Tab:MB-VLLs}, lower mass bounds for the vector-like leptons are summarized. 
The charge assignments are given in Table~\ref{Tab:VLLs}. 
Since the electrons and muons can be a clean signal at hadron colliders, 
the vector-like leptons that couple to the first two generation have been searched. 
If the vector-like lepton only decays to tau lepton modes, 
weaker experimental bounds are expected. 

The triplet vector-like lepton $N_1$, which appears in Type-III seesaw model, 
is analyzed by assuming the degeneracy among different charge components. 
The bound is relatively strong because of the large cross section through 
the $s$-channel $W$ exchange diagram~\cite{Aad:2015dha}. 
The bound on the singlet vector-like lepton $E_0$ is weaker, 
since it is produced only in a pair via $Z$ bosons and photons  
(the production cross section is much smaller than that of triplet vector-like leptons via $W$ bosons). 
This result is also applicable for the singly charged component of $E_1$. 
A bound for the Higgsino decaying to $H$ or $Z$ with a gravitino 
can be reinterpreted for the bound on the neutral component of $L_{1/2}$~\cite{Khachatryan:2014mma}. 
Assuming the 50\% branching fractions both for $H$ and $Z$ modes, 
a lower mass bound of 300GeV is estimated. 
The bound for the charged component comes from the LEP direct search~\cite{Achard:2001qw}.

\footnotetext[1]{Except for $<$129 GeV and 144--163 GeV.}
\footnotetext[2]{Except for $<$144 GeV and 153--160 GeV.}
\footnotetext[3]{Except for 401--419 GeV.}
}

\def\thesection{Appendix \Alph{section}}
\section{Renormalization group equations}\label{app:rge}
Here, we give the one-loop RGEs of our models. First, the one-loop RGEs of the SM gauge couplings are as follows:

\begin{align}
\frac{dg_{Y}}{dt}&=\frac{g_{Y}^{3}}{(4\pi)^{2}}\left(\frac{41}{6}+N_{f}N_{\psi}n_{\psi}\frac{4}{3}Y_{\psi}^{2}+N_{f}N_{\phi}n_{\phi}\frac{1}{3}Y_{\phi}^2\right),\nonumber
\\
\frac{dg_{2}}{dt}&=\frac{g_{2}^{3}}{(4\pi)^{2}}\left(-\frac{19}{6}+N_{f}N_{\psi}\frac{4}{3}S_{2}(F)+N_{f}N_{\phi}\frac{1}{3}S_{2}(S)\right),\nonumber
\\
\frac{dg_{3}}{dt}&=\frac{g_{3}^{3}}{(4\pi)^{2}}\left(-7+N_{f}n_{\psi}\frac{4}{3}S^{(c)}_{2}(F)+N_{f}n_{\phi}\frac{1}{3}S^{(c)}_{2}(S)\right).\nonumber
\end{align}
Here, $t:=\log \mu$ with $\mu$ being the renormalization scale, $Y_{\phi}$ $(Y_{\psi})$ is the $U(1)_{Y}$ hypercharge of $\phi$ $(\psi)$, $n_{\phi}$ $(n_{\psi})$ is the $SU(2)_{L}$ representation of $\phi$ $(\psi)$, $N_{\phi}$ $(N_{\psi})$ is the $ SU(3)_{c}$ representation of $\phi$ $ (\psi)$, $S_{2}(S)$ $(S_{2}(F))$ is the $SU(2)_{L}$ Dynkin index of $\phi$ $(\psi)$, and $S_{2}^{(c)}(S)$ $(S^{(c)}_{2}(F))$ is the $SU(3)_{c}$ Dynkin index of $\phi$ $(\psi)$.

Next, we summarize the one-loop RGEs of the scalar quartic couplings for each scalar model.
Note that it is convenient to use the Fierz identity in order to see the number of independent terms in the scalar potential, Eq.~\eqref{eq:pot1}.
The Fierz identities are given by
\begin{align}
&(t^a)^i_j (t^a)^k_l
=
{1\over2}\left(-{1\over2}\delta^i_j \delta^k_l + \delta^i_l \delta^k_j \right),  \quad 
\{t^a,t^b\}^i_j \{t^a,t^b\}^k_l={3\over4}\delta^i_j \delta^k_l, 
\quad \text{for } SU(2)\,\, {\bf 2}, \nonumber
\\
&\{t^a,t^b\}^i_j \{t^a,t^b\}^k_l
=
A\delta^i_j \delta^k_l + (8-A)\delta^i_l \delta^k_j + (-6+A) (t^a)^i_j (t^a)^k_l + (4-A) (t^a)^i_l (t^a)^k_j, \quad \text{for } SU(2)\,\, {\bf 3}, \nonumber
\\
&(T^a)^i_j (T^a)^k_l
=
{1\over2}\left(-{1\over3}\delta^i_j \delta^k_l + \delta^i_l \delta^k_j \right), 
\quad
\{T^a,T^b\}^i_j \{T^a,T^b\}^k_l
={11\over18}\delta^i_j \delta^k_l+{5\over6} \delta^i_l \delta^k_j
\quad \text{for } SU(3)\,\, {\bf 3}, \nonumber
\\
&\{T^a,T^b\}^i_j \{T^a,T^b\}^k_l
=
A\delta^i_j \delta^k_l + \left({80\over9}-A\right) \delta^i_l \delta^k_j + \left(-{17\over3}+{3\over2}A\right) (T^a)^i_j (T^a)^k_l + \left({22\over3}-{3\over2}A\right) (T^a)^i_l (T^a)^k_j,  \nonumber\\
&\h{14cm}\text{for } SU(3)\,\, {\bf 6}, \nonumber
\\
&\{T^a,T^b\}^i_j \{T^a,T^b\}^k_l
=
3\delta^i_j \delta^k_l + 3 \delta^i_l \delta^k_j+3 \delta^i_k \delta^j_l - (T^a)^i_j (T^a)^k_l + 2 (T^a)^i_l (T^a)^k_j,  \quad \text{for } SU(3)\,\, {\bf 8} ,\nonumber
\\
&\text{Tr}(T^aT^bT^cT^d)
={3\over4}\delta^{ab} \delta^{cd} + {3\over4} \delta^{bc} \delta^{ad}+{3\over4} \delta^{ac} \delta^{bd} +{1\over2} (T^e)^a_c (T^e)^b_d + (T^e)^a_d (T^e)^b_c,  \quad \text{for } SU(3)\,\, {\bf 8}, \nonumber
\end{align}
where $A$ is a free parameter.
Among the various scalar quartic couplings, we can write the beta functions of $\lambda, \lambda_X, \kappa_{H\phi}, \kappa'_{H\phi}, \kappa_{HX}, \kappa_{\phi X}$ generally as follows: \footnote{
Regarding the terms including $\lambda_\text{tr}$ and $\lambda'_{\text{ad}}$, we assume $N_f=1$.
}

\aln{
&\frac{d\lambda}{dt}
=
\frac{1}{16\pi^{2}}
\bigg(
	{1\over4}S_2(S) N_\phi N_f \kappa_{H\phi}'^2+n_\phi N_\phi N_f \kappa _{H\phi}^2
	-9 g_2^2 \lambda -3 \lambda  g_Y^2
	+\frac{3}{4} g_2^2 g_Y^2
	 +\frac{3g_Y^4}{8}+\frac{9 g_2^4}{8}+\frac{\kappa _{\text{HX}}^2}{2}+24 \lambda ^2
	 \nonumber\\
	 &\h{14cm}+12\lambda  y_t^2-6 y_t^4\bigg),
	 \nonumber\\
&\frac{d\lambda_X}{dt}
=
\frac{1}{16\pi^{2}}
\bigg(
	3\lambda_X^2+12\kappa_{HX}^2+6n_\phi N_\phi N_f \kappa_{\phi X}^2
\bigg),
	 \nonumber\\
&\frac{d\kappa_{H\phi}}{dt}
=
\frac{1}{16\pi^{2}}
\bigg(
	6y_t^2\kappa_{H\phi} + 4 \kappa_{H\phi}^2 + C_2(S) \kappa_{H\phi}'^2
	+12\kappa_{H\phi}\lambda + \kappa_{\phi X}\kappa_{HX}
	+4C_2(S)\kappa_{H\phi}\lambda''_\phi+4C_2^{(c)}(S)\kappa_{H\phi}\lambda'_\phi
	\nonumber\\
	&\h{2cm}
	+4C_2(S)C_2^{(c)}(S)\kappa_{H\phi}\lambda'''_\phi
	+(4N_f n_\phi N_\phi+4)\kappa_{H\phi}\lambda_\phi + 4 (n_\phi+ N_\phi) \kappa_{H\phi} \lambda_\text{tr}
	+3Y_\phi^2 g_Y^4 + 3 C_2(S) g_2^4
	\nonumber\\
	&\h{2cm}
	-\left({9\over2}+6C_2(S)\right)\kappa_{H\phi} g_2^2
	-\left({3\over2}+6Y_\phi^2\right)\kappa_{H\phi} g_Y^2
	-6C_2^{(c)}(S)\kappa_{H\phi} g_3^2
	+4(n_\phi+1)\kappa_{H\phi}\lambda'_{\text{ad}}
\bigg),
	 \nonumber}
\aln{&\frac{d\kappa'_{H\phi}}{dt}
=
\frac{1}{16\pi^{2}}
\bigg(
	6y_t^2\kappa'_{H\phi} + 8 \kappa_{H\phi} \kappa'_{H\phi}
	+4\kappa'_{H\phi}\lambda +4\kappa'_{H\phi}\lambda_\phi 
	+4\left(C_2(S)-1+N_\phi S_2(S) N_f \right)\kappa'_{H\phi}\lambda''_\phi
	\nonumber\\
	&\h{2cm}	
	+4C_2^{(c)}(S)\kappa'_{H\phi}\lambda'_\phi
	+4(C_2(S)-1)C_2^{(c)}(S)\kappa'_{H\phi}\lambda'''_\phi+ 4 N_\phi \kappa'_{H\phi} \lambda_{\text{tr}}+4 \kappa'_{H\phi}\lambda'_{\text{ad}}
	\nonumber\\
	&\h{2cm}
	-\left({9\over2}+6C_2(S)\right)\kappa'_{H\phi} g_2^2
	-\left({3\over2}+6Y_\phi^2\right)\kappa'_{H\phi} g_Y^2
	-6C_2^{(c)}(S)\kappa'_{H\phi} g_3^2
	+12Y_\phi g_Y^2 g_2^2
\bigg),
	 \nonumber\\
&\frac{d\kappa_{HX}}{dt}
=
\frac{1}{16\pi^{2}}
\bigg(
	\kappa_{HX}(12\lambda+\lambda_X+4\kappa_{HX}+6y_t^2-{3\over2}g_Y^2-{9\over2}g_2^2)
	+2n_\phi N_\phi N_f \kappa_{\phi H}\kappa_{\phi X}\bigg),
	\nonumber\\
&\frac{d\kappa_{\phi X}}{dt}
=
\frac{1}{16\pi^{2}}
\bigg(
	4 \kappa_{\phi X}^2 + \kappa_{\phi X}\lambda_X +4 \kappa_{HX} \kappa_{H\phi }
	+4C_2(S)\kappa_{\phi X}\lambda''_\phi+4C_2^{(c)}(S)\kappa_{\phi X}\lambda'_\phi+
	4C_2(S)C_2^{(c)}(S)\kappa_{\phi X}\lambda'''_\phi
	\nonumber\\
	&
	\h{2cm}+(4N_f n_\phi N_\phi+4)\kappa_{\phi X}\lambda_\phi + 4 (n_\phi+ N_\phi) \kappa_{\phi X} \lambda_\text{tr}
	-6C_2(S)\kappa_{\phi X} g_2^2
	-6Y_\phi^2\kappa_{\phi X} g_Y^2
	-6C_2^{(c)}(S)\kappa_{\phi X} g_3^2
	\nonumber\\
	&
	\h{2cm}
	+4(n_\phi+1)\kappa_{\phi X}\lambda'_{\text{ad}}
\bigg).\nonumber
 }
In the following, we show the RGEs of other scalar quartic couplings for each scalar model. Notice that we neglect such a coupling that is not induced at one loop level if we put it zero at the weak scale, and that the $U(1)_{Y}$ hypercharge $Y_{\phi}$ is kept as a free parameter.

\begin{itemize}\item {\large $(\mathbf{1},\mathbf{1},Y_{\phi})$}
\end{itemize}
\aln{
&\frac{d\lambda_{\phi}}{dt}=\frac{1}{16\pi^{2}}\left(-12 \lambda_{\phi} g_Y^2 Y_{\phi}^2+(4N_f+16)\lambda_{\phi}^2 +\frac{\kappa
   _{\phi X}^2}{2}+6 g_Y^4 Y_{\phi}^4+2 \kappa _{H\phi}^2\right),\nonumber  
   }

\begin{itemize}\item {\large $(\mathbf{1},\mathbf{2},Y_{\phi})$}\footnote{$\lambda_\phi''$ term exists only if $N_f\geq2$.}
\end{itemize}

\aln{
&\frac{d\lambda_{\phi}}{dt}=\frac{1}{16\pi^{2}}\bigg(-12 \lambda _{\phi} g_Y^2 Y_{\phi}^2-9g_{2}^{2}\lambda_\phi+ (8N_f+16) \lambda _{\phi}^2+\frac{\kappa
   _{\phi X}^2}{2} + {3\over2}\lambda_\phi''^2+6\lambda_\phi\lambda_\phi''
   +3 g_2^2 g_Y^2 Y_{\phi}^2\delta_{N_f,1}+6 g_Y^4
   Y_{\phi}^4
   \nonumber\\
   &\h{13cm}+\frac{9 g_2^4}{8}+\frac{1}{8} \kappa{'}_{H\phi}^2\delta_{N_f,1}+2 \kappa_{H\phi}^2\bigg),
   \nonumber\\  
&
\frac{d\lambda''_{\phi}}{dt}=\frac{1}{16\pi^{2}}
					\bigg(
					12Y_\phi^2 g_Y^2 g_2^2 + {1\over2} \kappa_{H\phi}'^2 + (2N_f-2) \lambda_\phi''^2 - 12 Y_\phi^2 g_Y^2 \lambda_\phi''- 9 g_2^2 \lambda_\phi''+ 24 \lambda_\phi \lambda_\phi''				
					\bigg),
\nonumber
\\     
   \nonumber}

\begin{itemize}\item {\large $(\mathbf{1},\mathbf{3},Y_{\phi})$}\footnote{
Notice that $\lambda_{\phi2}'$ term exists only if $N_f\geq2$, and that $\lambda_{\text{ad}}$ is not induced at one loop level.
}
\end{itemize}
\aln{
&\frac{d\lambda_{\phi2}'}{dt}=\frac{1}{16\pi^{2}}\left(-24 g_2^2 \lambda '_{\text{$\phi $2}}-12 Y^2 g_Y^2 \lambda '_{\text{$\phi
   $2}}+\frac{3 g_2^4}{2}+24 \lambda _{\phi } \lambda '_{\text{$\phi $2}}-8
   \lambda ''_{\phi } \lambda '_{\text{$\phi $2}}+2 \lambda{''}_{\phi
   }^2+(16N_f+152) \lambda{'}_{\text{$\phi $2}}^2
   \right),\nonumber\\ 
&\frac{d\lambda_{\phi}''}{dt}=\frac{1}{16\pi^{2}}\bigg(-24 g_2^2 \lambda{''}_{\phi}-12 g_Y^2 Y_{\phi}^2 \lambda{''}_{\phi}+(8N_f+8- 4\delta_{N_f,1}) \lambda{''}_\phi^2+24 \lambda _{\phi} \lambda{''}_{\phi}+12 g_2^2 g_Y^2 Y_{\phi}^2-3
   g_2^4\delta_{N_{f},1}
   \nonumber\\ 
   &\h{13cm}
   +\frac{1}{2} \kappa{'}_{H\phi}^2+128\lambda'_{\phi2}\lambda{''}_{\phi}
   \bigg),\nonumber\\ 
&\frac{d\lambda_{\phi}}{dt}=\frac{1}{16\pi^{2}}\bigg(-24 g_2^2 \lambda _{\phi}-12 \lambda _{\phi} g_Y^2 Y_{\phi}^2+16 \lambda{''}_\phi^2\delta_{N_f,1}+16  \lambda _{\phi} \lambda{''}_{\phi}+(12N_{f}+16) \lambda _{\phi}^2+\frac{\kappa
   _{\phi X}^2}{2}+6 g_Y^4 Y_{\phi}^4+12 g_2^4\delta_{N_f,1}\nonumber\\
  &\h{8cm} +2 \kappa _{H\phi}^2+(128N_f+96)\lambda_{\phi}\lambda_{\phi2}'+256N_f\lambda_{\phi2}^{'2}+128\lambda''_{\phi}\lambda'_{\phi2}
  \bigg),\nonumber\\     
   \nonumber}

\begin{itemize}\item {\large $(\mathbf{3}(\mathbf{3^{*}}),\mathbf{1},Y_{\phi})$}\footnote{
$\lambda_{\phi}'$ term exists only if $N_f\geq2$.}

\end{itemize}

\aln{
&\frac{d\lambda_{\phi}}{dt}=
   \frac{1}{16\pi^{2}}\bigg(-16 g_3^2 \lambda _{\phi}-12 g_Y^2 \lambda _{\phi} Y_{\phi}^2+4 g_3^2 g_Y^2
   Y_{\phi}^2\delta_{N_f,1}+6 g_Y^4 Y_{\phi}^4+\left({4\over3}+{5\over6}\delta_{N_f,1}\right) g_3^4
   +2 \kappa
   _{H\phi}^2+(12 N_{f}+16) \lambda _{\phi}^2
   \nonumber\\
   &\h{13cm}+\frac{\kappa _{\phi X}^2}{2}+{32\over3}\lambda_\phi\lambda_\phi'+{16\over9}\lambda_\phi'^2   
   \bigg),\nonumber\\  
  & 
 \frac{d\lambda_{\phi}'}{dt}
 =\frac{1}{16\pi^{2}}\left(-16g_3^2\lambda_\phi' -12g_Y^2\lambda_\phi'Y_\phi^2 +12 Y_\phi^2 g_Y^2 g_3^2 +{5\over2}g_3^4
   					+24\lambda_\phi\lambda_\phi'+2(N_f+1)\lambda_\phi'^2
 \right),\nonumber
  }

\begin{itemize}\item {\large $(\mathbf{3}(\mathbf{3^{*}}),\mathbf{2},Y_{\phi})$}\footnote{
$\lambda_\phi''$ and $\lambda_\phi'''$ terms exist only if $N_f\geq2$.
}
\end{itemize}

\aln{
&\frac{d\lambda_{\phi}}{dt}=\frac{1}{16\pi^{2}}\bigg(-9 g_2^2 \lambda _{\phi}-16 g_3^2 \lambda _{\phi}-12 g_Y^2
   Y_{\phi}^2 \lambda _{\phi}+(24N_f+16) \lambda _{\phi}^2
 -g_2^2 g_Y^2 Y_{\phi}^2 \delta_{N_f,1}+ 6 g_Y^4 Y_{\phi}^4+\frac{9 g_2^4}{8}
   +\frac{8}{3} g_3^2 g_2^2 \delta_{N_f,1}
     \nonumber\\
   &\h{4cm}
    + \frac{4 g_3^4}{3}
   +2
   \kappa _{H\phi}^2+\frac{\kappa _{\phi X}^2}{2}
   +{32\over3}\lambda_\phi \lambda_\phi'+6\lambda_\phi\lambda_\phi'' + 8 \lambda_\phi \lambda_\phi''' +{3\over2}\lambda_\phi''^2
   +{16\over9}\lambda_\phi'^2+{1\over3}\lambda_\phi'''^2
   \bigg),\nonumber\\  
  & \frac{d\lambda^{'}_{\phi}}{dt}=\frac{1}{16\pi^{2}}\bigg(-16 g_3^2 \lambda '_{\phi}-9 g_2^2 \lambda '_{\phi}-12
   g_Y^2 Y_{\phi}^2 \lambda '_{\phi}+(4 N_{f} + 2)\lambda{'}_{\phi}^2+ 24 \lambda _{\phi} \lambda'_{\phi}
   +{3}\lambda_\phi''\lambda_\phi''' - \lambda_\phi\lambda_\phi'''
   \nonumber\\
   &\h{4cm}
   +12 g_3^2 g_Y^2 Y_{\phi}^2+12 g_2^2 g_Y^2
   Y_{\phi}^2\delta_{N_f,1}+\frac{5 g_3^4}{2}-5 g_2^2 g_3^2\delta_{N_f,1}+{5\over8}\lambda_\phi'''^2
   +6\lambda_\phi'\lambda_\phi''-\lambda_{\phi}'\lambda_{\phi}'''
  \bigg),\nonumber\\ 
  & 
  \frac{d\lambda^{''}_{\phi}}{dt}=\frac{1}{16\pi^{2}}\bigg(-16 g_3^2 \lambda ''_{\phi}-9 g_2^2 \lambda ''_{\phi}-12
   g_Y^2 Y_{\phi}^2 \lambda ''_{\phi}+(6 N_{f}-2) \lambda{''}_{\phi}^2+24 \lambda _{\phi} \lambda''_{\phi}+12 g_2^2 g_Y^2
   Y_{\phi}^2\nonumber\\
   &\h{10cm}
   +\frac{1}{2}\kappa{'}_{H\phi}^2+{32\over3}\lambda_\phi'\lambda_\phi'' - {8\over3} \lambda_\phi'' \lambda_\phi'''
   +\frac{32}{9}\lambda_\phi' \lambda_\phi'''
   \bigg),
  \nonumber}
 \aln{ & 
  \frac{d\lambda^{'''}_{\phi}}{dt}=\frac{1}{16\pi^{2}}\bigg(-16 g_3^2 \lambda '''_{\phi}-9 g_2^2 \lambda '''_{\phi}-12
   g_Y^2 Y_{\phi}^2 \lambda '''_{\phi}+12 g_2^2 g_3^2
   +\left(N_f+{10\over3}\right)\lambda_\phi'''^2 + \frac{16}{3}\lambda_\phi' \lambda_\phi'''
   +16\lambda_{\phi}'\lambda_{\phi}''
   \nonumber\\
   &\h{12cm}+24\lambda_{\phi}\lambda_{\phi}'''-2\lambda_{\phi}''\lambda_{\phi}'''
   \bigg),
  \nonumber
   }

\begin{itemize}\item {\large $(\mathbf{3} (\mathbf{3^{*}}),\mathbf{3},Y_{\phi}),\ N_f=1$ \footnote{
$\lambda_{\text{ad}}$ is not induced at one loop level.
}}
\end{itemize}

\aln{
&\frac{d\lambda_{\phi}}{dt}=\frac{1}{16\pi^{2}}\bigg(-24 g_2^2 \lambda _{\phi}-16 g_3^2 \lambda _{\phi}-12 g_Y^2 \lambda _{\phi} Y_{\phi}^2+6
   g_Y^4 Y_{\phi}^4+8 g_2^4+4 g_3^2 g_2^2
   +\frac{4 g_3^4}{3}+2 \kappa
   _{H\phi}^2+16 \lambda _{\phi}
   \lambda ''_{\phi}\nonumber\\
   &\h{5cm}
   +{16\over9}\lambda_\phi'^2
   +{32\over3}\lambda_\phi''^2
   +\frac{32}{3} \lambda_\phi \lambda '_{\phi}
   +\frac{16}{3}\lambda_{\phi}'\lambda_{\phi}'' 
   +52 \lambda
   _\phi^2+\frac{\kappa_{\phi X}^2}{2}
   \bigg),\nonumber\\  
  &\frac{d\lambda^{'}_{\phi}}{dt}=\frac{1}{16\pi^{2}}\bigg(-24 g_2^2 \lambda '_{\phi}-16 g_3^2 \lambda '_{\phi}-12 g_Y^2 Y_{\phi}^2 \lambda
   '_\phi+12 g_3^2 g_Y^2 Y_{\phi}^2+12 g_2^4-12 g_3^2 g_2^2 
   +\frac{5
   g_3^4}{2}
   +8\lambda_\phi'^2 
    \nonumber\\
   &\h{12cm}+ 16\lambda_\phi''^2
   +
   24 \lambda _{\phi} \lambda '_{\phi}+16\lambda_\phi'\lambda_\phi''
   \bigg),\nonumber\\ 
  & \frac{d\lambda_{\phi}''}{dt}=\frac{1}{16\pi^{2}}\bigg(-24 g_2^2 \lambda ''_{\phi}-16 g_3^2 \lambda ''_{\phi}-12 g_Y^2 Y_{\phi}^2 \lambda
   ''_\phi+12 g_2^2 g_Y^2 Y_{\phi}^2-3 g_2^4+4 g_3^2 g_2^2 
   +
   28\lambda_\phi''^2
   +\frac{1}{2}
   \kappa{'}_{H\phi}^2
    \nonumber\\
   &\h{14cm}
   +24
   \lambda _{\phi} \lambda ''_{\phi}
   +\frac{32}{3} \lambda '_{\phi} \lambda ''_{\phi}
   \bigg),\nonumber
   }

\begin{itemize}\item {\large $(\mathbf{6^{*}},\mathbf{1},Y_{\phi}),\  N_f=1$}
\end{itemize}

\aln{
&\frac{d\lambda_{\phi}}{dt}=\frac{1}{16\pi^{2}}\bigg(-40 g_3^2 \lambda _{\phi}-12 \lambda _{\phi} g_Y^2 Y_{\phi}^2+\frac{160}{9}
   \lambda{'}_\phi^2+\frac{80}{3} \lambda _{\phi} \lambda '_{\phi}+40 \lambda
   _\phi^2+\frac{\kappa _{\phi X}^2}{2}+6 g_Y^4 Y_{\phi}^4+\frac{40
   g_3^4}{3}+2 \kappa _{H\phi}^2\bigg),\nonumber\\  
   &\frac{d\lambda_{\phi}'}{dt}=\frac{1}{16\pi^{2}}\bigg(-40 g_3^2 \lambda '_{\phi}-12 g_Y^2 Y_{\phi}^2 \lambda '_{\phi}+28 \lambda{'}_\phi^2+24 \lambda _{\phi} \lambda '_{\phi}+12 g_3^2 g_Y^2
   Y_{\phi}^2+\frac{5 g_3^4}{2}\bigg),\nonumber
   }

\begin{itemize}\item {\large $(\mathbf{6^{*}},\mathbf{3},Y_{\phi}),\  N_f=1$ \footnote{\ $\lambda_{\text{ad}}$ is not induced at one loop level.}}
\end{itemize}

\aln{
&\frac{d\lambda_{\text{tr}}}{dt}=\frac{1}{16\pi^{2}}\bigg(-32 \lambda'''{}_{\phi } \lambda _{\text{tr}}+16 \lambda _{\text{tr}} \lambda '_{\phi
   }+24 \lambda _{\phi } \lambda _{\text{tr}}+36 \lambda
   _{\text{tr}}^2-24 g_2^2 \lambda _{\text{tr}}-40 g_3^2 \lambda _{\text{tr}}-12
   g_Y^2 \lambda _{\text{tr}} Y_{\phi }^2   +6g_2^4+6g_3^4
   \nonumber\\
   &
   \h{10cm}+8\lambda_\phi'^2+8\lambda_\phi''^2
   +{176\over9} \lambda{'''}_{\phi }^2
   -16\lambda_\phi' \lambda_\phi'''
   -{32\over3}\lambda_\phi'' \lambda_\phi'''
   \bigg),\nonumber}
\aln{&\frac{d\lambda'''_{\phi}}{dt}=\frac{1}{16\pi^{2}}\bigg(36 \lambda{'''}_{\phi }^2+24 \lambda _{\phi} \lambda'''
   {}_{\phi }+\frac{64}{3} \lambda'''{}_{\phi } \lambda '_{\phi
   }+16 \lambda '_{\phi } \lambda ''_{\phi }+8 \lambda'''{}_{\phi }
   \lambda _{\text{tr}}-24 g_2^2 \lambda'''{}_{\phi }-40 g_3^2 \lambda'''
   {}_{\phi }
   -12 g_Y^2 \lambda'''{}_{\phi } Y_{\phi }^2
   \nonumber\\
   &\h{15cm}+12g_3^2 g_2^2
   \bigg),
   \nonumber\\
&\frac{d\lambda''_{\phi}}{dt}=\frac{1}{16\pi^{2}}\bigg(-{80\over9} \lambda{'''}_{\phi }^2+52 \lambda
   {''}_{\phi }^2+24 \lambda _{\phi } \lambda ''_{\phi }+ \frac{80}{3} \lambda'''{}_{\phi } \lambda
   ''_{\phi }+\frac{320}{9} \lambda'''{}_{\phi } \lambda '_{\phi
   }+\frac{80}{3} \lambda '_{\phi } \lambda ''_{\phi }\nonumber\\
   &\h{2.5cm}+\frac{80}{3} \lambda'''
   {}_{\phi } \lambda _{\text{tr}}+64 \lambda _{\text{tr}} \lambda
   ''_{\phi }-24 g_2^2 \lambda ''_{\phi }-40 g_3^2 \lambda ''_{\phi }-12
   g_Y^2 Y_{\phi }^2 \lambda ''_{\phi }
   +12 g_2^2 g_Y^2 Y_{\phi }^2-3 g_2^4+\frac{1}{2} \kappa{'}_{\text{H$\phi $}}^2
   \bigg),\nonumber\\  
&\frac{d\lambda'_{\phi}}{dt}=\frac{1}{16\pi^{2}}\bigg(\frac{20}{3} \lambda{'''}_{\phi }^2+48 \lambda
   {'}_{\phi }^2+24 \lambda _{\phi } \lambda '_{\phi }+\frac{88}{3}
   \lambda'''{}_{\phi } \lambda '_{\phi }+16 \lambda '_{\phi } \lambda
   ''_{\phi }+32\lambda ''_{\phi }\lambda '''_{\phi }+16 \lambda'''{}_{\phi }
   \lambda _{\text{tr}}+40 \lambda _{\text{tr}} \lambda '_{\phi } \nonumber\\
   &\h{8cm}
   -40 g_3^2
   \lambda '_{\phi }-24 g_2^2 \lambda '_{\phi }-12 g_Y^2 Y_{\phi }^2 \lambda
   '_{\phi }+12 g_3^2 g_Y^2 Y_{\phi }^2+\frac{5 g_3^4}{2}
   \bigg),\nonumber\\  
  & \frac{d\lambda_{\phi}}{dt}=\frac{1}{16\pi^{2}}
  \bigg(16\lambda{'''}_{\phi }^2+\frac{160}{3} \lambda _{\phi }
   \lambda'''{}_{\phi }+8 \lambda{''}_{\phi }^2+16 
   \lambda _{\phi } \lambda ''_{\phi }+\frac{88}{9} \lambda
   {'}_{\phi }^2+\frac{80}{3} \lambda _{\phi } \lambda '_{\phi
   }+\frac{32}{3} \lambda'''{}_{\phi } \lambda ''_{\phi } +16 \lambda'''
   {}_{\phi } \lambda '_{\phi }+88 \lambda _{\phi }^2
   \nonumber\\
   &\h{2cm}
  +32
   \lambda'''{}_{\phi } \lambda _{\text{tr}}+16 \lambda _{\text{tr}}
   \lambda ''_{\phi }+\frac{32}{3} \lambda _{\text{tr}} \lambda '_{\phi }
   +72
   \lambda _{\phi } \lambda _{\text{tr}}+12 \lambda _{\text{tr}}^2  -24
   g_2^2 \lambda _{\phi }-40 g_3^2 \lambda _{\phi }-12 g_Y^2 \lambda _{\phi }
   Y_{\phi }^2
   +6 g_Y^4 Y_{\phi }^4
   \nonumber\\
   &\h{12cm}+6 g_2^4
   +\frac{22
   g_3^4}{3}+2 \kappa _{\text{H$\phi $}}^2+\frac{\kappa _{\text{$\phi
   $X}}^2}{2}
   \bigg),\nonumber 
   }

\begin{itemize}\item {\large $(\mathbf{8},\mathbf{2},Y_{\phi}), \ N_f=1$ }
\end{itemize}
\aln{
&\frac{d\lambda_\phi}{dt}
=\frac{1}{16\pi^{2}}\bigg(
					-36  g_3^2 \lambda_\phi - 9 g_2^2 \lambda_\phi -12 Y_\phi^2 g_Y^2 \lambda_\phi	
					+{9\over8}g_2^4 + \frac{27}{4} g_3^4 + 6 g_Y^4 Y_\phi^4
					+2\kappa_{H\phi}^2+{1\over2}\kappa_{\phi X}^2
					\nonumber\\
					&\h{2cm}
					+80\lambda_\phi^2 
					+ 9 \lambda_\phi'^2
					+ {3\over2}\lambda_\phi''^2
					+ 6 \lambda_\phi \lambda_\phi''
					+ 18 \lambda_\phi \lambda_\phi'''
					+ \frac{9}{2} \lambda_\phi' \lambda_\phi'''
					+ \frac{27}{16}\lambda_\phi'''^2
					+24\lambda_{\phi} \lambda_{\phi}'
					+12\lambda_{\text{ad}}'^2
					+24\lambda_{\text{ad}}'\lambda_\phi
   \bigg),\nonumber\\ 
&\frac{d\lambda'_\phi}{dt}
=\frac{1}{16\pi^{2}}\bigg(
					-36  g_3^2 \lambda'_\phi - 9 g_2^2 \lambda'_\phi -12 Y_\phi^2 g_Y^2 \lambda'_\phi
					 + 12 Y_\phi^2 g_Y^2 g_3^2
					+36\lambda_\phi'^2
					+24\lambda_\phi \lambda_\phi'
					+6 \lambda_\phi' \lambda_\phi''
					\nonumber\\
					&\h{10cm}
					+3 \lambda_\phi'' \lambda_\phi'''
					+ 12\lambda_\phi' \lambda_\phi'''
					-24\lambda_{\text{ad}}'\lambda_{\phi}'
					-6\lambda_{\text{ad}}'\lambda_{\phi}'''
   \bigg),\nonumber\\ 
&\frac{d\lambda''_\phi}{dt}
=\frac{1}{16\pi^{2}}\bigg(
					-36  g_3^2 \lambda''_\phi - 9 g_2^2 \lambda''_\phi -12 Y_\phi^2 g_Y^2 \lambda''_\phi	
					+ 9 g_3^4 + 12 Y_\phi^2 g_Y^2 g_2^2	
					+{1\over2}(\kappa'_{H\phi})^2
					+14\lambda_\phi''^2
					+12\lambda_\phi'^2
					\nonumber\\
					&\h{6cm}
					+24 \lambda_\phi \lambda_\phi''
					+24 \lambda_\phi' \lambda_\phi''
					-6 \lambda_\phi'' \lambda_\phi'''
					+ 6 \lambda_\phi' \lambda_\phi'''
					+ \frac{9}{4} \lambda_\phi'''^2+
					32\lambda_{\text{ad}}'^2
					+8\lambda_{\text{ad}}'\lambda_{\phi}''
   \bigg),\nonumber\\ 
&\frac{d\lambda'''_\phi}{dt}
=\frac{1}{16\pi^{2}}\bigg(
					-36  g_3^2 \lambda'''_\phi - 9 g_2^2 \lambda'''_\phi 
					-12 Y_\phi^2 g_Y^2 \lambda'''_\phi
					+ 6 g_3^4 +12 g_2^2 g_3^2
					+8 \lambda_\phi'^2
					+24 \lambda_\phi \lambda_\phi'''
					+16 \lambda_\phi' \lambda_\phi''
					\nonumber\\
					&\h{10cm}
					+4 \lambda_\phi' \lambda_\phi'''
					+ \frac{15}{2}\lambda_\phi'''^2
					-2\lambda_{\phi}''\lambda_{\phi}'''
					+16\lambda_{\text{ad}}\lambda_{\phi}'''
   \bigg),\nonumber}
\aln{ &
 \frac{d\lambda_{\text{ad}}'}{dt}
=\frac{1}{16\pi^{2}}  \bigg(-36 g_3^2 \lambda_{\text{ad}}'-9 g_2^2 \lambda_{\text{ad}}'+\frac{9 g_3^4}{2}
   -12 \lambda_{\text{ad}}'g_Y^2 Y_{\phi }^2
   +\frac{9}{8} \lambda{'''}_{\phi }^2
   -6 \lambda_{\text{ad}}' \lambda '''{}_{\phi }
   +10\lambda_{\text{ad}}' \lambda ''_{\phi }
   +6 \lambda {'}_{\phi }^2
   -24\lambda_{\text{ad}}' \lambda '_{\phi }
   \nonumber\\
   &\h{10cm}
   +3 \lambda '''{}_{\phi } \lambda'_{\phi}       
   +24 \lambda_{\text{ad}}' \lambda _{\phi}
   +32 \lambda_{\text{ad}}'^2\bigg).
   \nonumber
   }

\end{document}